\documentclass[10pt,aps,prb,twocolumn,nofootinbib]{revtex4-1}

\usepackage{amssymb,amsmath}
\usepackage{float}
\usepackage{graphicx}
\usepackage{color}
\usepackage{diagbox}
\usepackage{hyperref}
\pdfoutput=1
\begin{document}

\title{A Systematic Approach to Generating Accurate Neural Network Potentials: the Case of Carbon}

\author{Yusuf Shaidu}
\affiliation{Scuola Internazionale Superiore di Studi Avanzati, Trieste, Italy.}
\affiliation{The Abdus Salam International Centre for Theoretical Physics, Trieste, Italy.}

\author{Emine K\"{u}\c{c}\"{u}kbenli}
\affiliation{John A. Paulson School of Engineering and Applied Sciences, Harvard University, Cambridge, Massachusetts 02138, USA}
\affiliation{Scuola Internazionale Superiore di Studi Avanzati, Trieste, Italy.}

\author{Ruggero Lot}
\affiliation{Scuola Internazionale Superiore di Studi Avanzati, Trieste, Italy.}

\author{Franco Pellegrini}
\affiliation{Laboratoire de Physique de l'\'{E}cole normale sup\'{e}rieure, ENS, Universit\'{e} PSL, CNRS, Sorbonne Universit\'{e}, Universit\'{e} de Paris, F-75005 Paris, France}

\author{Efthimios Kaxiras}
\affiliation{John A. Paulson School of Engineering and Applied Sciences, Harvard University, Cambridge, Massachusetts 02138, USA}
\affiliation{Department of Physics, Harvard University, Cambridge, MA 02138, USA}

\author{Stefano  de Gironcoli}
\affiliation{Scuola Internazionale Superiore di Studi Avanzati, Trieste, Italy.}
\affiliation{CNR-IOM DEMOCRITOS, Istituto Officina dei Materiali, Trieste, Italy.}

\date{\today}

\begin{abstract}
Availability of affordable and widely applicable interatomic potentials is the key needed to unlock the riches of modern materials modelling. 
Artificial neural network based approaches for generating potentials are promising;
however neural network training requires large amounts of data, sampled adequately from an often unknown potential energy surface. 
Here we propose a self-consistent approach that is based on crystal structure prediction formalism 
and is guided by unsupervised data analysis,
to construct an accurate, inexpensive and transferable artificial neural network potential.  
Using this approach, we construct an interatomic potential for Carbon 
and demonstrate its ability to reproduce first principles results on elastic and vibrational properties for diamond, graphite and graphene, as well as energy ordering and structural properties of a wide range of crystalline and amorphous phases. 
\end{abstract}

\maketitle

\section{Introduction}

The state of the art theoretical framework for computing material properties of crystals at their ground state is Density Functional Theory (DFT)~\cite{HK1964,KS1965}. 
DFT allows to describe the total energy as a functional of electron density, E[$\rho$], for a given atomic configuration $\{\mathbf{R}\}$, by taking advantage of the conjugate relationship between the electrostatic potential of the nuclei $V(\{\mathbf{R}\})$, and the ground state electron density $\rho$. 
By solving the expensive quantum mechanical equations that result from this definition for electrons, DFT outlines a path to determine the total energy, the forces on each atom, the stress due to crystal structure, and several other ground-state properties of materials. 
Yet the cost of solving the quantum mechanical equations, as well as having to work with the extensive electronic wavefunctions and density, hinders the application of this method to systems beyond a few thousands of atoms. 

A way to reduce the computational cost lies in the realization that the same conjugate relationship between $\rho$ and $V$ guarantees that 
a functional exists which maps the electrostatic potential of the nuclei to the total energy, 
hence it is possible to describe ground state properties as a functional of the positions of atoms in the structure, without having to work explicitly with the electron density. 
Yet, the exact form of such a functional is unknown. 
One approach to approximate this unknown functional is using artificial neural networks (ANNs). 
ANNs and in general machine learning techniques, have been shown to yield reasonably accurate functional approximations for a wide range of applications, 
and have already been adopted with success to some material science problems~\cite{Chandrasekaran2019, Xie2018, Onat2018, Kolsbjerg2018, Cooper2020, THOMPSON2015316, Zong2018, HIMANEN2019, Nyshadham2019, Kostiuchenko2019,Deng2019,Schmidt2019_npj,Zubatyukeaav6490}.  

ANNs can be seen as an attractive alternative to the classical approach for constructing interatomic interaction models (also known as force fields) where physical intuition is used to fix the form of the approximate functional for $E[V(\{\mathbf{R}\})]$. 
While physically meaningful forms can describe the interatomic interaction in a compact way, with only few parameters to be fitted, the rigidity of the functional form reduces the predictive power of this method in exploratory studies. 
In particular, for highly polymorphic materials such as Carbon, where several different bonding types and structures exist, 
the lack of transferability of a model from one structure to another results in many different interaction models, each with a limited applicability. 
For example, among the several empirical force fields for carbon, the non-reactive, short range, bond-order-based Tersoff~\cite{Tersoff1988} model can describe dense $sp^3$ carbon structures 
while a highly parametric reactive force field (ReaxFF)~\cite{vanDuin2001} that explicitly includes long-range van der Waals interactions and Coulomb energy through charge equilibration scheme~\cite{Rappe1991} is needed for structures with $sp^2$ hybridization.
Furthermore, even though these empirical force fields give a qualitative understanding of materials\ properties, they are quantitatively inaccurate when compared to both {\it ab initio} methods and experiments~\cite{Khaliullin2010,Koukaras2015,Deringer_amorphous_2017, Wen2019}.

Interatomic interaction models based on ANNs do not have a fixed functional form beyond the network architecture, 
and their parameters are fitted to vast amounts of {\it ab initio} quantum mechanical data in the hope of assimilating the physics of the system into the parametrization. 
Hence the transferability {\it restraint} of classical force fields, that is due to their rigid form, 
is traded for a transferability {\it challenge} in the case of neural networks due to the (lack of) variety and completeness in the training set. 
To address this challenge of generating truly transferable ANN interatomic interaction models, training data must be obtained from an efficient and thorough sampling of the potential energy landscape. 
Such sampling of the very rugged and high dimensional landscape with {\it ab initio} electronic structure tools is a formidable challenge. 

In this work we integrate evolutionary algorithm with molecular dynamics in a self-consistent manner to sample the potential energy landscape and obtain data with high variability. 
Moreover, for reliable materials modelling, it is crucial to have indicators that signal when the limit of transferability is crossed.
We address this aspect of ANN models by studying the relationship between data variability and transferability of the trained network via unsupervised data analysis.
We demonstrate the performance of the approach highlighted above on the challenging example of crystalline and amorphous Carbon structures. 

This study is a continuation of similar efforts in the literature: 
The first ANN interaction model for elemental carbon was developed in 2010 by Khaliullin {\it et al.}~\cite{Khaliullin2010} to study graphite-diamond co-existence. 
The network was trained on an adaptive training set, where the starting configurations were manually selected from randomly distorted graphite and  diamond phases, relaxed under a range of external pressures (from -10 to 200 GPa) at zero temperature. 
Then, configurations for new training data were obtained using this model in finite temperature molecular dynamics (MD) simulations, 
which in turn were used to refine the network, until a self-consistency was reached in the prediction error on the new structures. 
More recently in 2019, a hybrid  model, where an ANN potential for the short-range interaction is supplemented with a theoretically motivated analytical term to model long-range dispersion, 
has been developed in order to address the properties of monolayer and multilayer graphene, with encouraging results~\cite{Wen2019}. 
As we will demonstrate in this work, ANN models such as these, built on data sampled solely from a limited part of the potential energy landscape can however be highly non-transferable. 
This transferability challenge for Carbon has been observed with kernel-based machine learning models as well. 
\begin{figure}[t]
    \centering
    \includegraphics[scale=0.23]{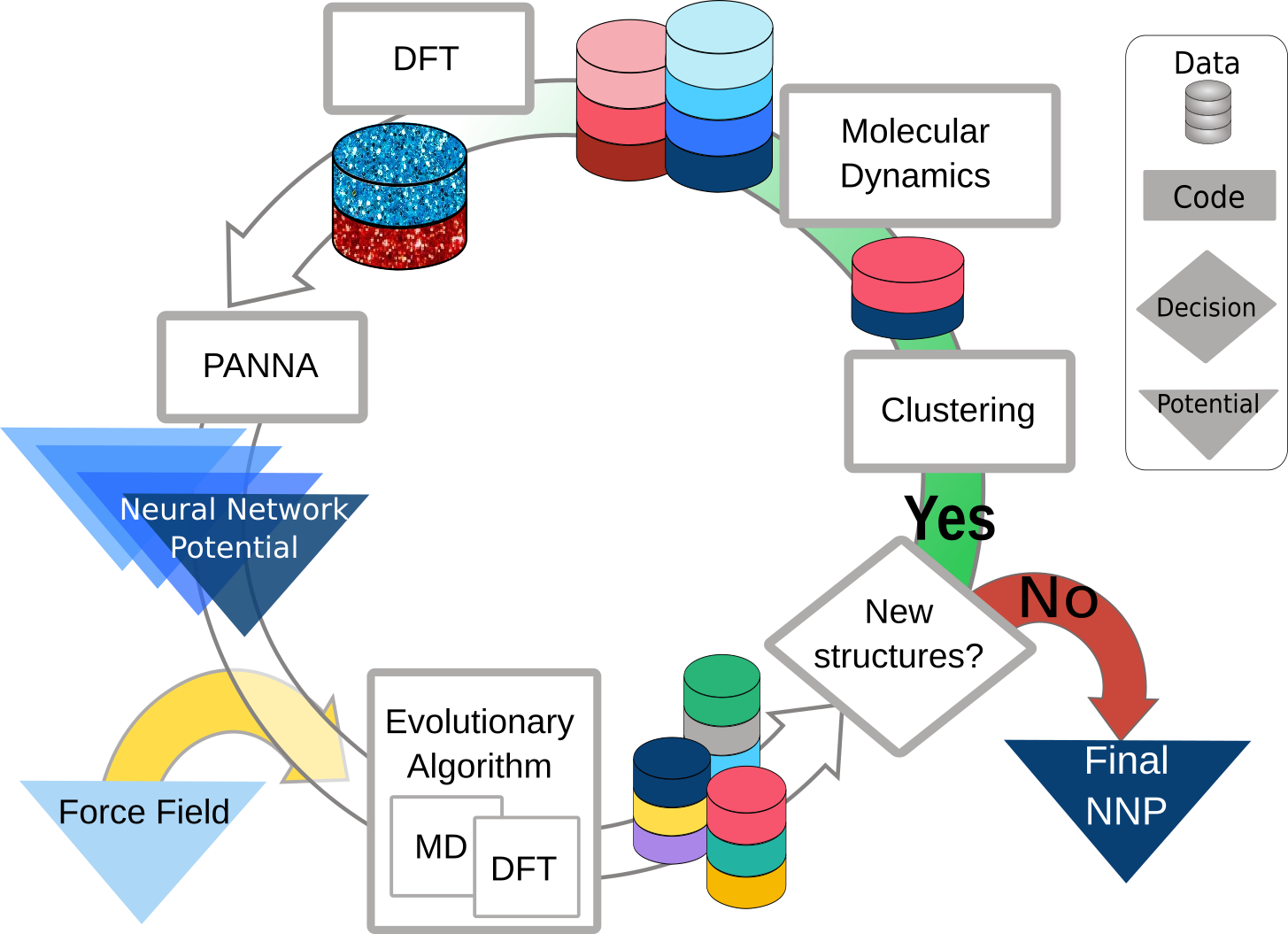}
    \caption{
    Sketch of the self-consistent scheme to generate an accurate and transferable neural network potential. The initial step to start the process (yellow arrow) can be performed with a classical force field as shown here, or any comprehensive dataset of structures such as the ones in Aflowlib~\cite{CURTAROLO2012}, Materials Genome Initiative~\cite{DEPABLO2014} or Nomad~\cite{NOMAD18}  repositories can be used to generate the first neural network potential model (blue triangle) to be refined through the self-consistent cycle. Once an initial potential model is chosen, Evolutionary Algorithm enables a diverse set of structures to be sampled. The following clustering-based pruning of structures further ensures that no single polymorph biases the dataset, i.e. at each step only novel structures (red and blue disks for the particular step highlighted above) are to be considered, further refined, and added to the dataset. The subsequent MD simulations sample the potential energy surface of each polymorph. Finally DFT calculations performed on a subset of MD-sampled structures are added to the {\it ab initio} dataset obtained thus far. The {\it ab initio} dataset augmented this way is then used to train the next neural network potential model (a darker blue triangle), starting the next cycle of the self-consistent scheme until no new structures are found by the Evolutionary Algorithm.}
    \label{fig:Workflow}
\end{figure}

\begin{figure*}[ht!]
    \centering
    \includegraphics[width=\textwidth]{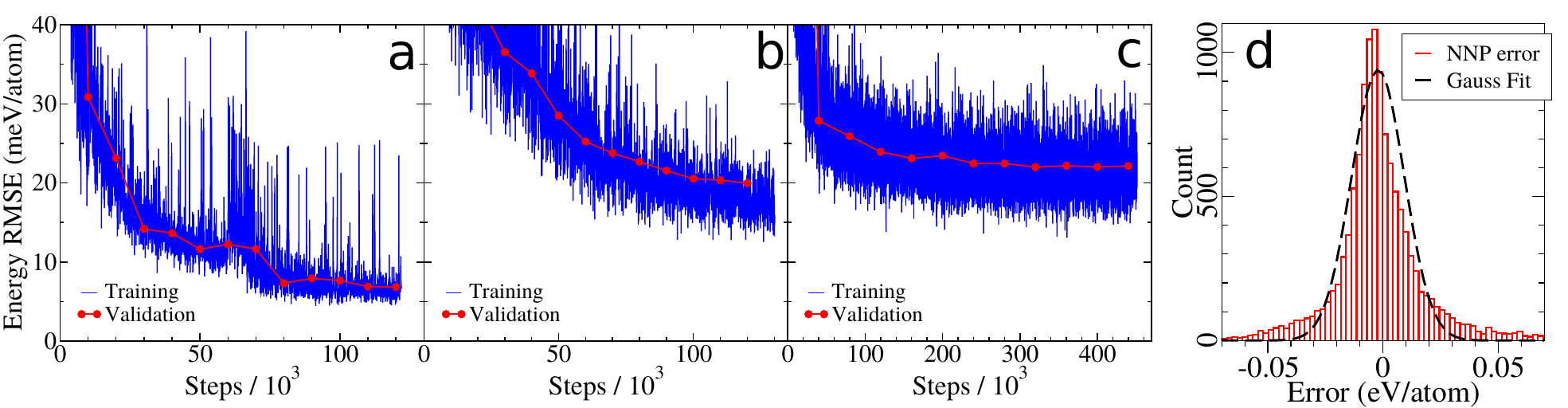}
    \caption{Training and validation error in energy: the evolution of error in energy on training and validation set for potentials trained at first ({\bf a}), second ({\bf b}) and third ({\bf c}) iteration of the self-consistent cycle. The blue lines are the RMSE on a given batch of 128 configurations during training. The networks are evaluated during training on all the validation set of sizes $\approx$3000, $\approx$5200 and $\approx$12000 configurations for first, second and third iterations respectively (red dots with lines as guide to eye). The final training and validation RMSE are reported in Table~\ref{tab:scf_rmse}.
    ({\bf d}) Error distribution for the validation dataset at third iteration. The black dashed line is a normalized Gaussian fit, resulting in an RMSE of $11.3$~meV, clearly failing to fit the fat tailed distribution.
    }\label{fig:train_validate}
 \end{figure*}  

 In 2017, a kernel-based model was constructed~\cite{Deringer_amorphous_2017} using data from MD melt-quench trajectories of liquid and amorphous Carbon, to study amorphous structures. 
Motivated from its non-optimal behavior on crystalline phases, authors developed another model with a specialized training data obtained via MD, for graphene~\cite{RowePatrick2018}. 
It is worthwhile to note that recently, a strategy combining kernel-based model generation with crystal structure prediction was suggested by Bernstein {\it et al.}~\cite{Bernstein2019}. 
Since computational cost for training or evaluation of a kernel-based model grows with the training set however, this approach is suitable for small scale configuration space sampling. 
In comparison, computational cost of neural networks are independent of the size of the training dataset, a feature that is exploited in the current study for accurate prediction of elastic and vibrational properties.

In this work we use a systematic approach to construct a highly flexible and transferable neural network potential (NNP) and demonstrate its application to the development of a general NNP for Carbon. We compare its performance with respect to other potential models previously optimized for specific phases and discuss the implications of our results for the trade-off between transferability and specialization.

\section{Results}\label{result}
\subsection{Self-Consistent Training and Validation}
The neural network potential is constructed following the self-consistent approach sketched in Fig.~\ref{fig:Workflow}. 
This recursive data-creating and fitting cycle starts with a trial force field (FF) which is used to generate an initial set of configurations via evolutionary algorithm. 
Evolutionary algorithms (EA) are commonly used in crystal structure prediction studies as they allow efficient sampling of the configuration space. 
Their success in thorough sampling is demonstrated by their ability to predict new crystal structures before the experimental observation~\cite{Ma2009, Bull2017}.  
As the exploration of the configuration space continues, a single point DFT calculation is performed on each distinct polymorph generated by EA. 
These structures are then clustered using a distance measure.
From each cluster, a representative example is manually 
selected and a classical molecular dynamics simulation at a given pressure and temperature range is performed. 
The additional MD simulation step allows the sampling of the whole neighborhood of the equilibrium configuration for each polymorph, resulting in accurate prediction of structural properties for every polymorph. 
The dataset obtained this way is used to train a neural network model. The trained NNP is then used for starting a new iteration of the self-consistent cycle.
This increases the training set diversity, by preventing the energetically favorable structures that are easily accessed by EA from dominating the whole training set. 
The iterative procedure highlighted above is repeated until no new structures are found. The methodological details of the self-consistent training used in this work is detailed in the Methods section.

\begin{table}[b]
\caption{Training and Validation RMSE: The training RMSE is the average over the batch RMSE of the last 2500 training steps, while the validation RMSE is evaluated over the entire validation set with the NNP obtained at the last training step. The energy (force) mean absolute error on the validation set are 4 (0.09), 12(0.12) and 14(0.16)\,meV/atom (eV/\AA{}) for first, second and third iteration respectively. 
}
\label{tab:scf_rmse}
\begin{ruledtabular}
\begin{tabular}{cl|rr|rr} 
No. of&Size of & \multicolumn{2}{c|}{Energy RMSE} & \multicolumn{2}{c}{Force RMSE} \\
iterations & data & \multicolumn{2}{c|}{(meV/atom)} & \multicolumn{2}{c}{(eV/\AA{})}\\
~ & ~ & train & validation & train & validation\\
\hline
1 & 15841 &  6.8 &  6.8 & 0.14 & 0.14\\
2 & 30815 & 17.1 & 20.0 & 0.19 & 0.22\\
3 & 60133 & 22.0 & 22.1 & 0.26 & 0.27\\
\end{tabular}
\end{ruledtabular}
\end{table}

The performance of an NNP at each self-consistent loop is evaluated during training via the validation scheme. Fig.~\ref{fig:train_validate} shows the evolution of NNP energy accuracy on the training and validation set as a function of training steps at each self-consistent iteration (panels {\bf a}-{\bf c}). 
The training root-mean-square error (RMSE) corresponds to the instantaneous RMSE computed on the elements of the batch considered at that training step while the validation RMSE is computed on all the configurations in the validation set. 
The RMSE on the validation set agrees with the training RMSE throughout the training, an indication that the model does not overfit to the training data set. The analysis of the force prediction error at different stages of training gives similar results and can be found in the Supplementary Material.
The increase in energy and force RMSE from iteration 1 to 3 is a result of the increase in the diversity of atomic environments. 
At each self-consistent iteration the diversity of the dataset increases as new structures are explored (see Table~\ref{tab:scf_rmse}), while the number of parameters of the network, therefore its capacity, is kept fixed. 
It is worth noting that the prediction error is not distributed according to a Gaussian distribution function but a fatter-tailed one (see Fig.~\ref{fig:train_validate}{\bf d}). 
Therefore, while the RMSE given here is a good measure to compare training and validation error with one another, it overestimates the average NNP prediction error in general.

\begin{figure}[b]
    \centering
    \includegraphics[width=0.5\textwidth]{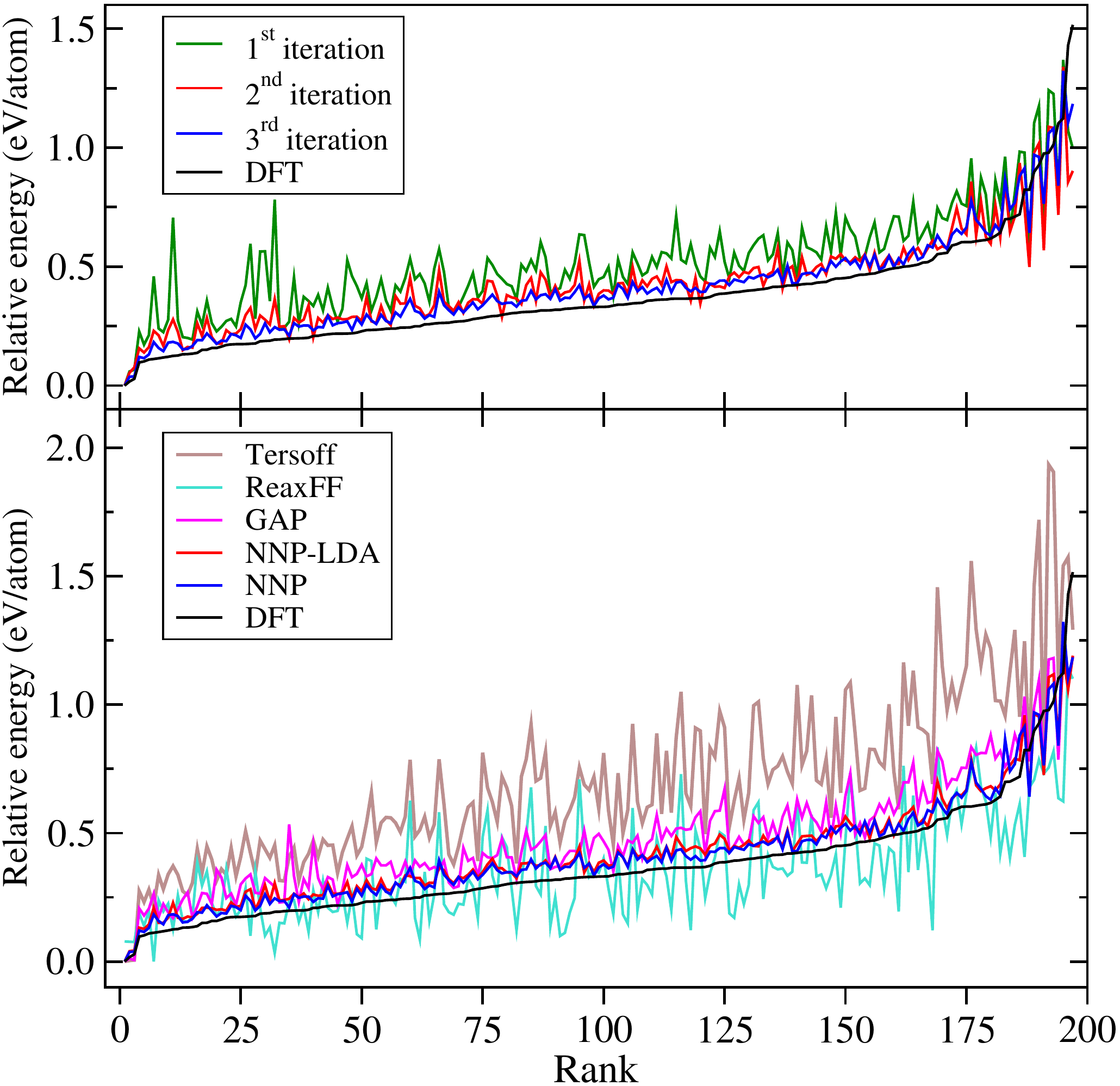}
    \caption{Energy ordering of the 197 distinct carbon structure reported in Ref.~\onlinecite{Deringer_structures_2017}. {\bf Top:} Performance of NNP at different iterations of the self-consistent cycle. {\bf Bottom:} Performance of GAP, reactive force field (ReaxFF)~\cite{Raju2015} and Tersoff~\cite{Tersoff1988} models compared to the final NNP model (\textit{blue line}). For comparison we train a new model with LDA exchange correlation functional, named as NNP-LDA (\textit{red line}). The neural network potentials for the two functionals overlap for majority of the structures.}
    \label{fig:energy_ordering}
\end{figure}
To demonstrate how the general accuracy of the NNPs is changing with each iteration, we check their performance on a dataset of 197 distinct carbon structures. 
These structures were obtained by Deringer and co-workers~\cite{Deringer_structures_2017} via random search of crystal structure of carbon with a Gaussian approximation potential (GAP) developed for liquid and amorphous carbon systems~\cite{Deringer_amorphous_2017} and are distributed online~\cite{WebHomeHomelibAtomsorg}. 
For consistency, their energies are re-calculated with the same DFT parameters as explained in Methods section.
Fig.~\ref{fig:energy_ordering} shows the energy ranking as predicted by NNP, GAP, Tersoff and ReaxFF. It can be seen that the NNP accuracy gets better with each iteration. 
The third iteration NNP accuracy agrees remarkably well with DFT results and performs better than all the other methods tested. 
It is noteworthy that the final NNP carries no signature of the ReaxFF used in the initial step to explore the configuration space. 
Both classical potentials, Tersoff and ReaxFF, perform very poorly compared to machine learnt ones, and the NNP outperforms GAP results published in Ref.~\onlinecite{Deringer_amorphous_2017,Deringer_structures_2017}, albeit GAP was fitted on
{\it ab initio} data obtained with LDA exchange correlation functional~\cite{Perdew1992}. 
For fair comparison, we train a new NNP, using the same training dataset obtained via the self consistent procedure, but with LDA functional. 
This potential, referred as NNP-LDA, performs similarly to the NNP highlighted in this work, and similarly outperforms all the other potentials. 
In the rest of the work, the results denoted with NNP refers to the potential that is trained with the rVV10 functional unless otherwise specified.

\begin{table*}[hbt!]
\begin{ruledtabular}
\centering
\caption{Elastic properties of Diamond, Graphite  and Graphene. 
For Diamond ({\bf top}) all machine learnt potentials reproduce their reference DFT lattice parameter with less than 1\% relative error. When the comparison extends to elastic constants and bulk modulus, however, only the NNP described in this work shows a consistently close agreement between DFT and the potential model, $<3\%$ relative error, the range of variation also observed between two different experiments. 
For Graphite ({\bf middle}) we report the lower bound for the bulk modulus using Reuss average, i.e. $ 1/B_0  \equiv  s_{11} + s_{22} + s_{33} + 2(s_{12} + s_{23} + s_{31})$. The robust intraplanar structural features of graphite is captured well by all machine learnt potentials while the weaker interplanar interaction and, in particular, elastic properties that couple the two, are more challenging to capture. This is true even for the hybrid potential hNN-Gr of Ref.\onlinecite{Wen2019} where the distance dependence of the long range interaction is manually set to $r^{-6}$ and the potential is tailor-fit to describe multi-graphene systems.
For Graphene ({\bf bottom}) the 2D elastic constants were computed with the normalized 3D stress as $\sigma_{2D}(\epsilon)= E(\epsilon)/A_0$ at a given strain of $\epsilon$, where $E(\epsilon)$ is the total energy at $\epsilon$ and $A_0=\sqrt{3}a^2/2$ is the area of graphene plane. $E$ is the Young modulus and $\nu$ is Poisson's ratio. The elastic constant is converted to bulk properties in GPa by dividing by the interlayer distance $c/2$ of graphite reported in Table~\ref{Elastic_Properties} or in respective experimental reference. Both machine learnt potentials similarly overestimate the Poisson's ratio and underestimate the Young's modulus with respect to their DFT references. Differences between DFT references are of similar magnitude as the differences between NNP and DFT in each case.}
\label{Elastic_Properties}
%
%
\begin{tabular}{l|rr r| rr r| rr r| r r| rr}
&\multicolumn{2}{c}{This work} & ~  &\multicolumn{2}{c}{Ref~[\onlinecite{Deringer_amorphous_2017}]} & ~ 
&\multicolumn{2}{c}{Ref~[\onlinecite{Khaliullin2010}]} & ~
& \multicolumn{1}{c}{Ref~[\onlinecite{Tersoff1988}]} & ~
& & \\
Diamond
&NNP &DFT &~ 
&GAP &DFT &~
&NNP &DFT &~
&Tersoff &~
&Exp[\onlinecite{McSkimin1972}]   
&Exp[\onlinecite{Zouboulis1998}] \\\hline 
 $a$  (\AA{})&3.576& 3.584& ~ &3.539&3.532& ~ &3.569&3.570& ~ & 3.566& &-   &3.567\\ 
 $B_0 $ (GPa)&431  &425  & ~  &438  &466  & ~ &434  &439  & ~ & 426  & &442 &445\\
$C_{11}$ (GPa)&1054 &1044 & ~  &1090 &1101 & ~ &1016 &1056 & ~ &1074  & &1079(5)&1080 \\
$C_{12}$ (GPa)&119  &116  & ~  &112  &148  & ~ &142  &130  & ~ &102   & &124(5) &127\\
$C_{44}$ (GPa)&542  &547  & ~  &594  &592  & ~ &580  &567  & ~ &641   & &578(2)&576\\
\end{tabular}
%
%
\begin{tabular}{l|rr r|rr r|rr r|r}
&\multicolumn{2}{c}{This work}&~ 
&\multicolumn{2}{c}{Ref~[\onlinecite{Khaliullin2010}]}&~ 
&\multicolumn{2}{c}{Ref~[\onlinecite{Wen2019}]}&~ 
& ~\\
Graphite &NNP  &DFT &~ & NNP  &DFT &~ &hNN-Gr$_x$ &DFT &~ & Experiment \\\hline
 $a$  (\AA{}) & 2.471 & 2.471 &~& 2.467 & 2.467 &~& 2.467 & 2.466 &~& 2.464$^b$,~~ 2.463$^c$\\
 $c$ (\AA{}) & 6.732 & 6.719 &~& 6.688 & 6.815 &~& 6.804 & 6.800 &~& 6.712$^b$,~~ 6.712$^c$\\
 $B_0 $  (GPa)  & 48  & 40  &~& 48  & 37 &~& - & - &~& 36(11)$^b$\\
$C_{11}$ (GPa) & 1053 & 1048 &~& 1080 & 1069  &~& 978 & 1080 &~& 1060(20)$^a$, 1109(16)$^b$\\
$C_{12}$ (GPa) & 197 & 182 &~& 179 & 162 &~& 177 & 162 &~& 180(20)$^a$,~~139(36)$^b$ \\
$C_{13}$ (GPa) & -23 & -5 &~& 0  & -4   &~& -67 & -5 &~& 15(5)$^a$,~~~~~~0(3)$^b$\\
$C_{33}$ (GPa) & 57 & 43  &~& 52  & 40   &~& 40 & 33&~& 37(10)$^a$,~~~~~39(7)$^b$\\
$C_{44}$ (GPa) &-5   &4   &~&7   &5    &~&1.79&3.36&~& 0.27$^a$,~~~~~~5(3)$^b$\\
$C_{66}$ (GPa) &428  &433 &~& -  &-    &~&-&-&~& 485(11)$^b$ 
\end{tabular}
%
%
\begin{tabular}{l|rr r| rr r|r r|r}
&\multicolumn{2}{c}{This work} &~
&\multicolumn{2}{c}{Ref~[\onlinecite{Wen2019}]}&~
&Ref~[\onlinecite{Tersoff1988}] &~
&~\\
Graphene &NNP &DFT &~&hNN-Gr$_x$ &DFT &~ &Tersoff &~&Experiment \\\hline
a (\AA{}) & 2.470 & 2.470 && 2.467 & 2.466 && 2.530 && 2.46$^d$\\
$\nu$       & 0.244 & 0.173 && 0.197 & 0.149 && -0.158 &&-\\
E (GPa)     & 967 & 1015 && 1021 & 1060 && 1216 && 1015(149)$^e$,~2400(400)$^f$ \\
$C_{11}$(GPa) & 1028 & 1047 && 1062 & 1084 && 1247 && - \\
$C_{12}$  (GPa)& 251 & 181 && 209 & 161 && -197 && -\\
\end{tabular}
\end{ruledtabular}
\begin{itemize}
   \item[]Experiments: $^a$Ref[\onlinecite{Seldin1970}] 
    $^b$Ref[\onlinecite{Bosak2007}] 
    $^c$Ref[\onlinecite{MOHR2007}]
    $^d$Ref[\onlinecite{Cooper12}]
    $^e$Ref[\onlinecite{Lee_Changgu2008}]
    $^f$Ref[\onlinecite{Jae_Ung2012}]
\end{itemize}
\end{table*}
\subsection{Structural and elastic properties}
In this section, we discuss the performance of the NNP on the structural and elastic properties of select Carbon polymorphs, namely, diamond, graphite and graphene (See Table \ref{Elastic_Properties}). 
The equilibrium lattice parameters are obtained by minimizing the total energy until the force components on each atom are lower than 26\,meV/\AA\,for both 
DFT and NNP simulations. 
We also include results obtained with Tersoff potential, as well as other DFT and machine learning studies in literature. 

In the case of diamond, all machine learning methods agree reasonably well with the DFT results they were trained with, both for the equilibrium volume and elastic constants. 
The largest deviation is seen in $C_{12}$ prediction with GAP potential with $24\%$ relative error. 
For all properties tested, the predictions of NNP of the current study is within a relative error of $5\%$ with respect to DFT. 
It should be noted that the variation between DFT studies employing different exchange correlation functionals are larger than the difference between machine learnt models and the DFT results they are trained to reproduce. 
Tersoff potential, although it predicts the equilibrium volume well, fails to predict the $C_{44}$.  

In the more challenging case of graphite, $C_{11}$ and $C_{12}$ relate to the in-plane elastic properties while $C_{33}$ probes the relationship between strain and stress between the planes which are held together by van der Waals interactions. 
$C_{13}$ and $C_{44}$ couple the strong in-plane interaction with the weak out-of-plane ones, namely $C_{13}$ can be seen as a measure of interlayer dilation upon layer compression, and $C_{44}$ as a measure of response to shear deformation. 
The performance of the NNP on prediction of graphite elastic constants are aligned with this overview: 
For all potentials reported in the middle panel of Table~\ref{Elastic_Properties}, in-plane lattice parameter and elastic constants are better predicted than the ones that relate to out of plane interaction, indicating that more data or better training is needed to describe these more delicate properties. 
Yet it is encouraging that the general purpose NNP of the current work performs at least as well as other NNPs from literature that were developed with a focus on van der Waals systems such as graphite and multilayer graphene. In section \ref{sec:EstimatingAccNNP} we discuss how focusing on particular system could further improve on these predictions. 

\subsection{Vibrational properties}
\begin{figure*}[ht!]
   \centering
        \includegraphics[scale=0.26]{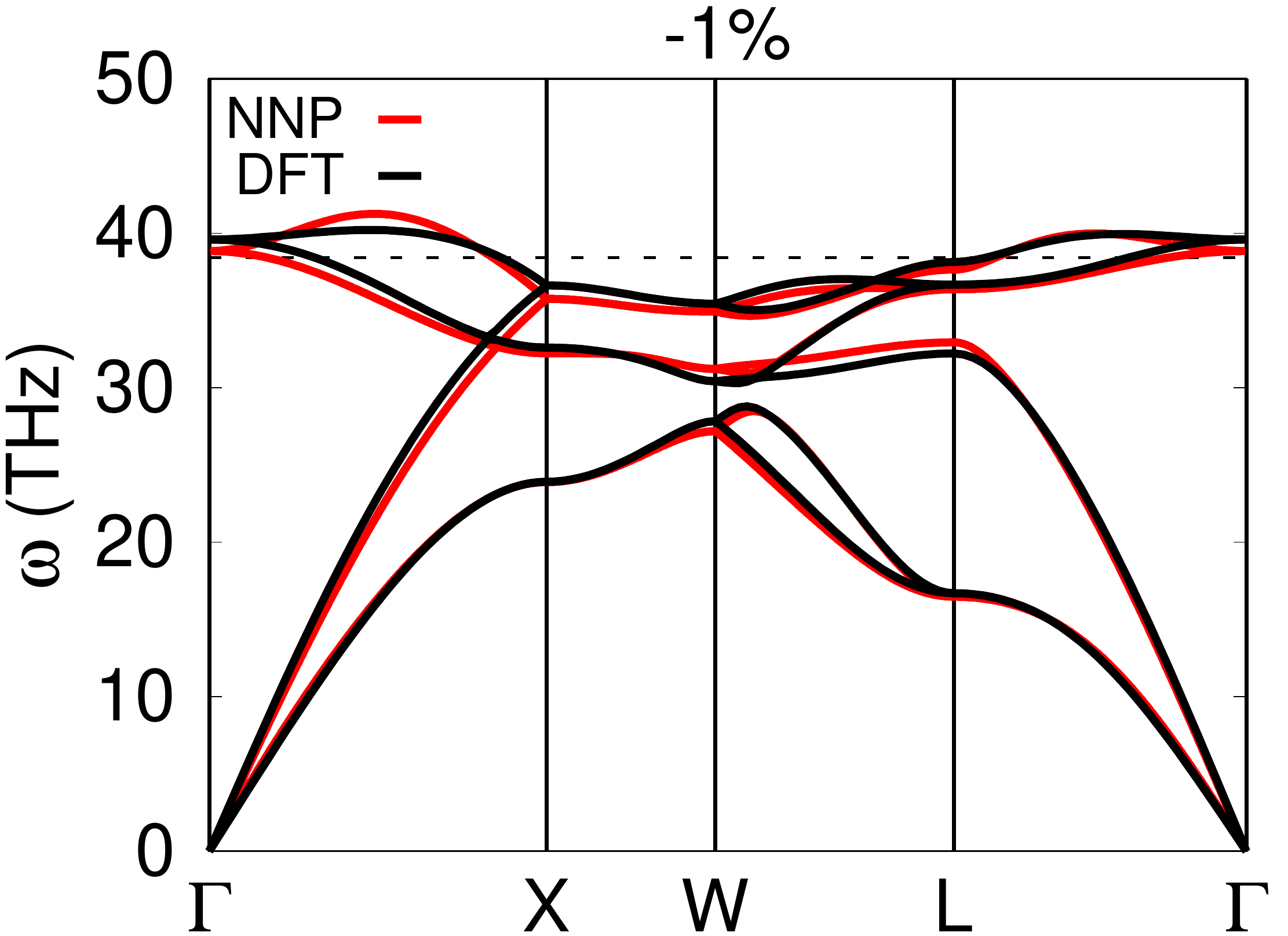}
        \includegraphics[scale=0.26]{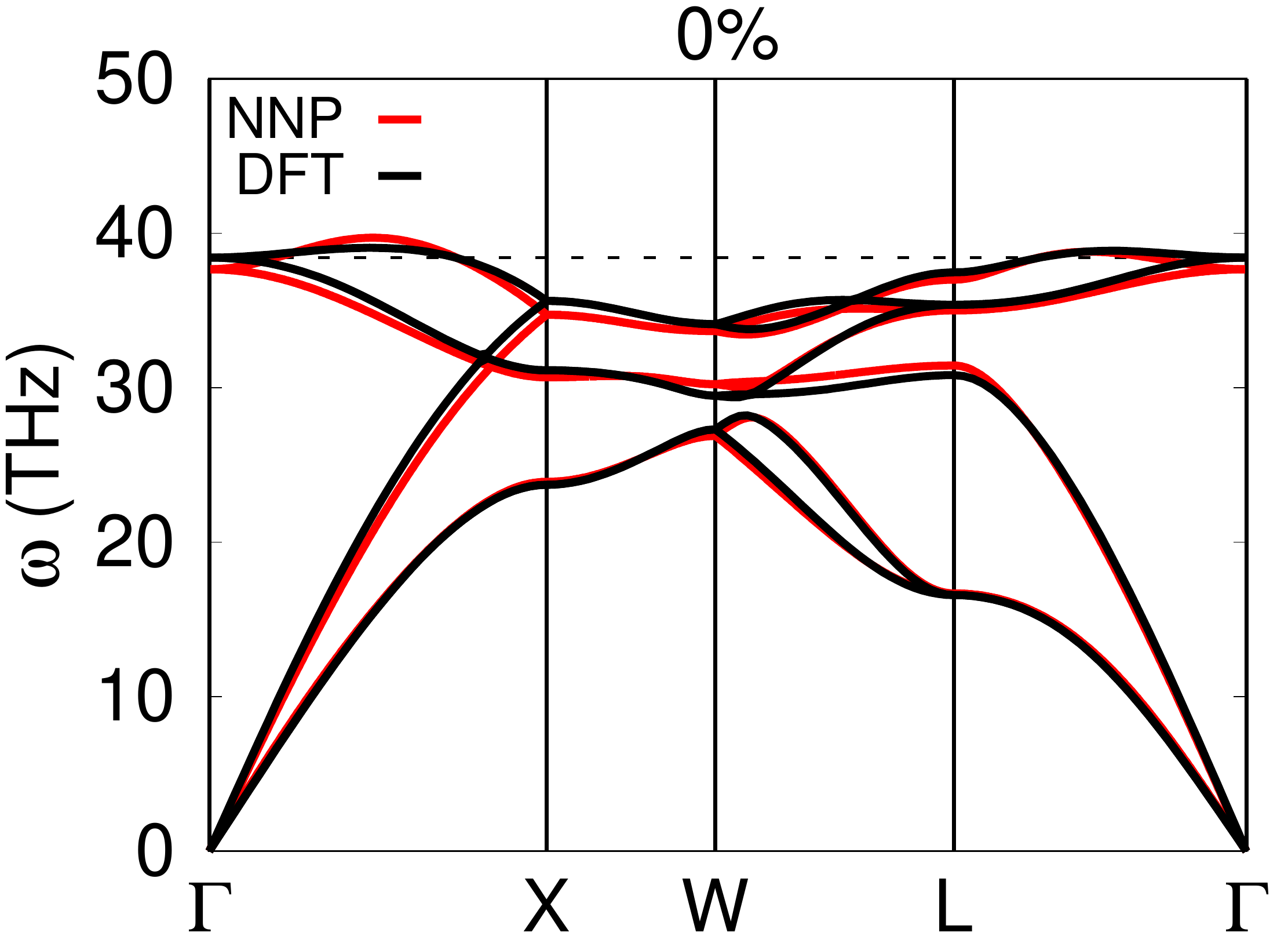}
        \includegraphics[scale=0.26]{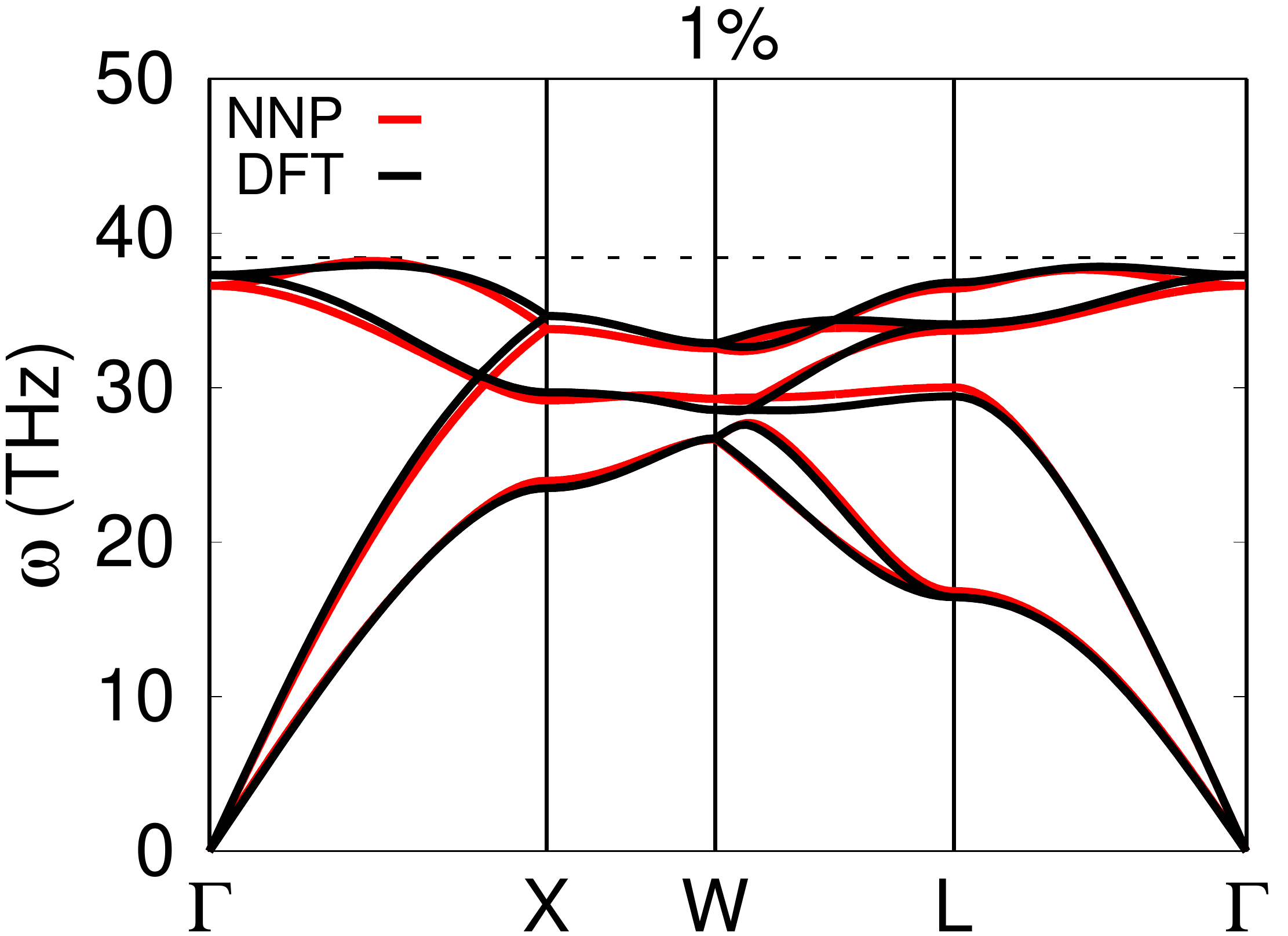}
  \par
        \includegraphics[scale=0.26]{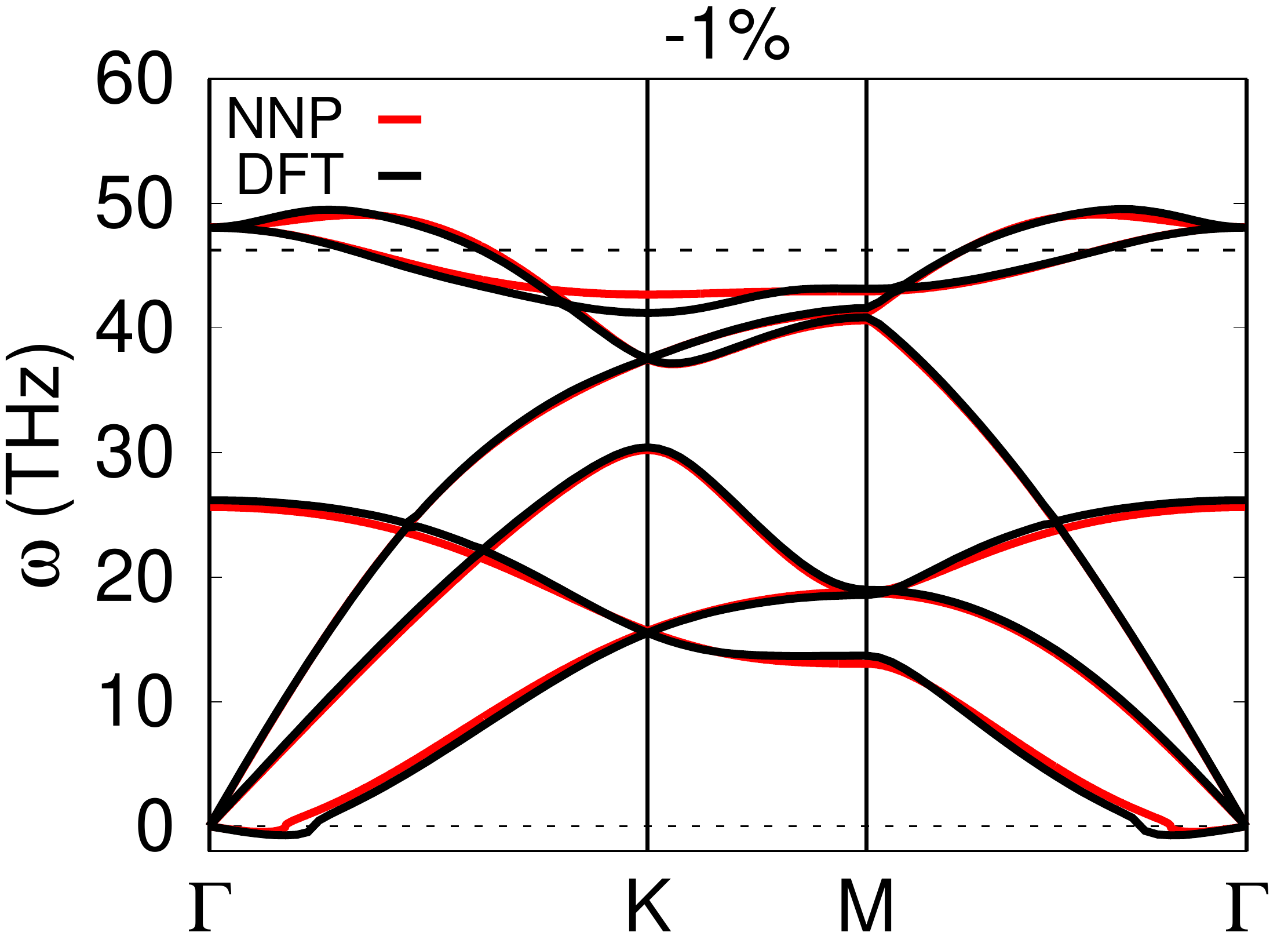}
        \includegraphics[scale=0.26]{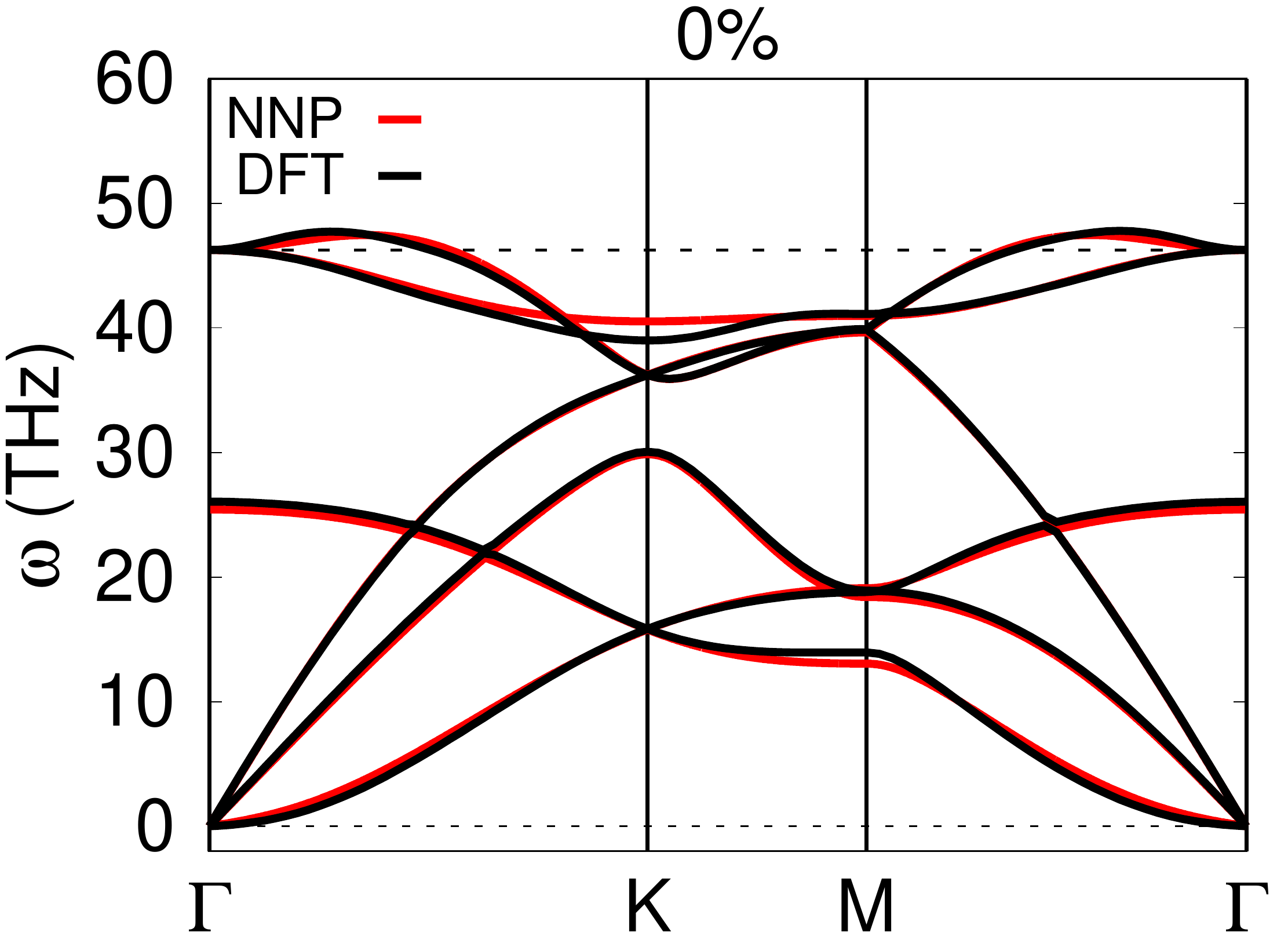}
        \includegraphics[scale=0.26]{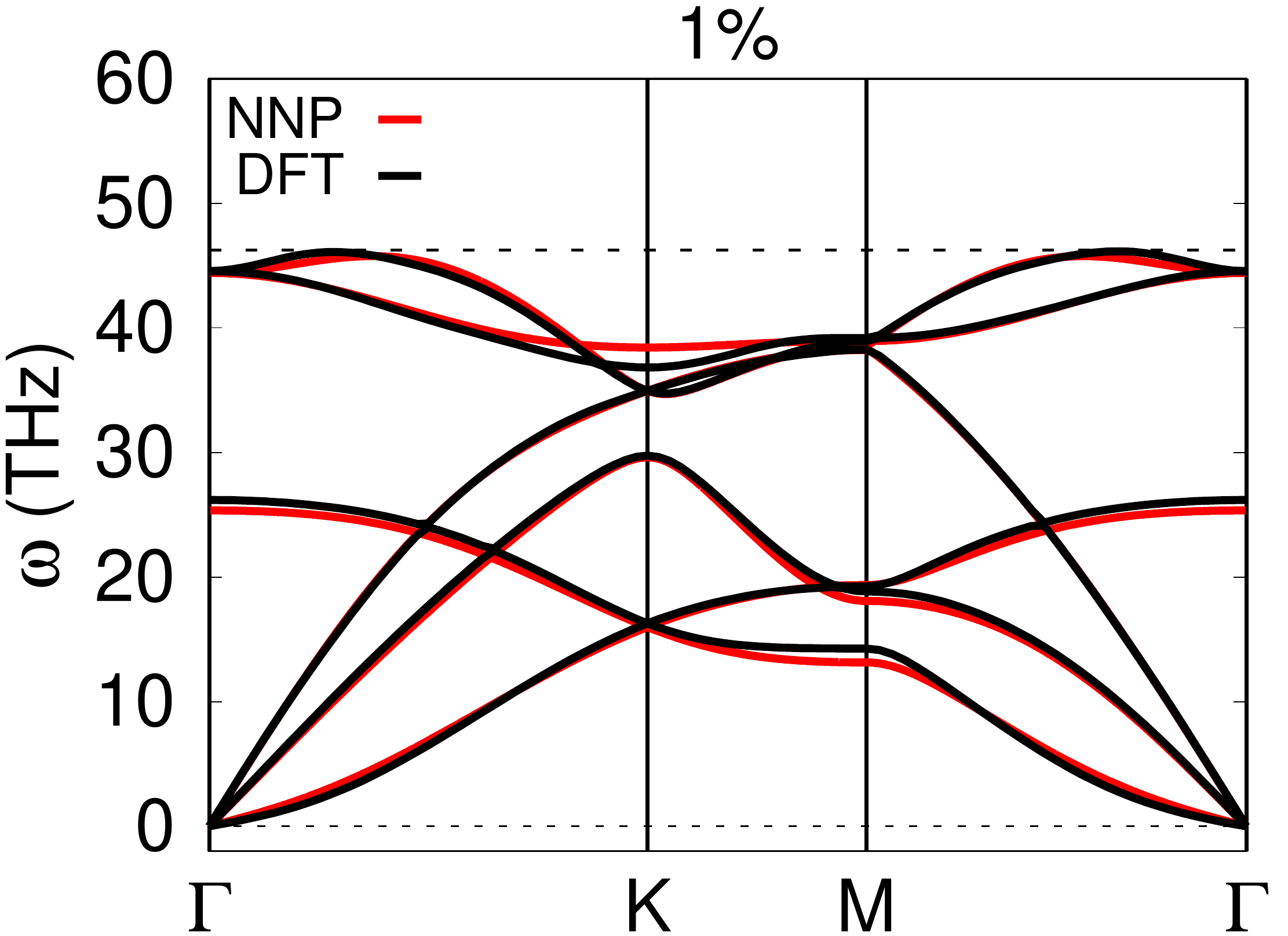}
    \caption{Phonon dispersion of diamond ({\bf top}) and graphene ({\bf bottom}) along the high symmetry lines. The value at the top of each graph represent the percentage of expansion (positive) or compression (negative) of the lattice parameter. The black dotted line is the maximum frequency in THz at the $\Gamma$ point at equilibrium lattice parameter.}
    \label{fig:diamond_phonon}
\end{figure*}

\begin{figure}[b!]
   \centering
        \includegraphics[width=0.33\textwidth]{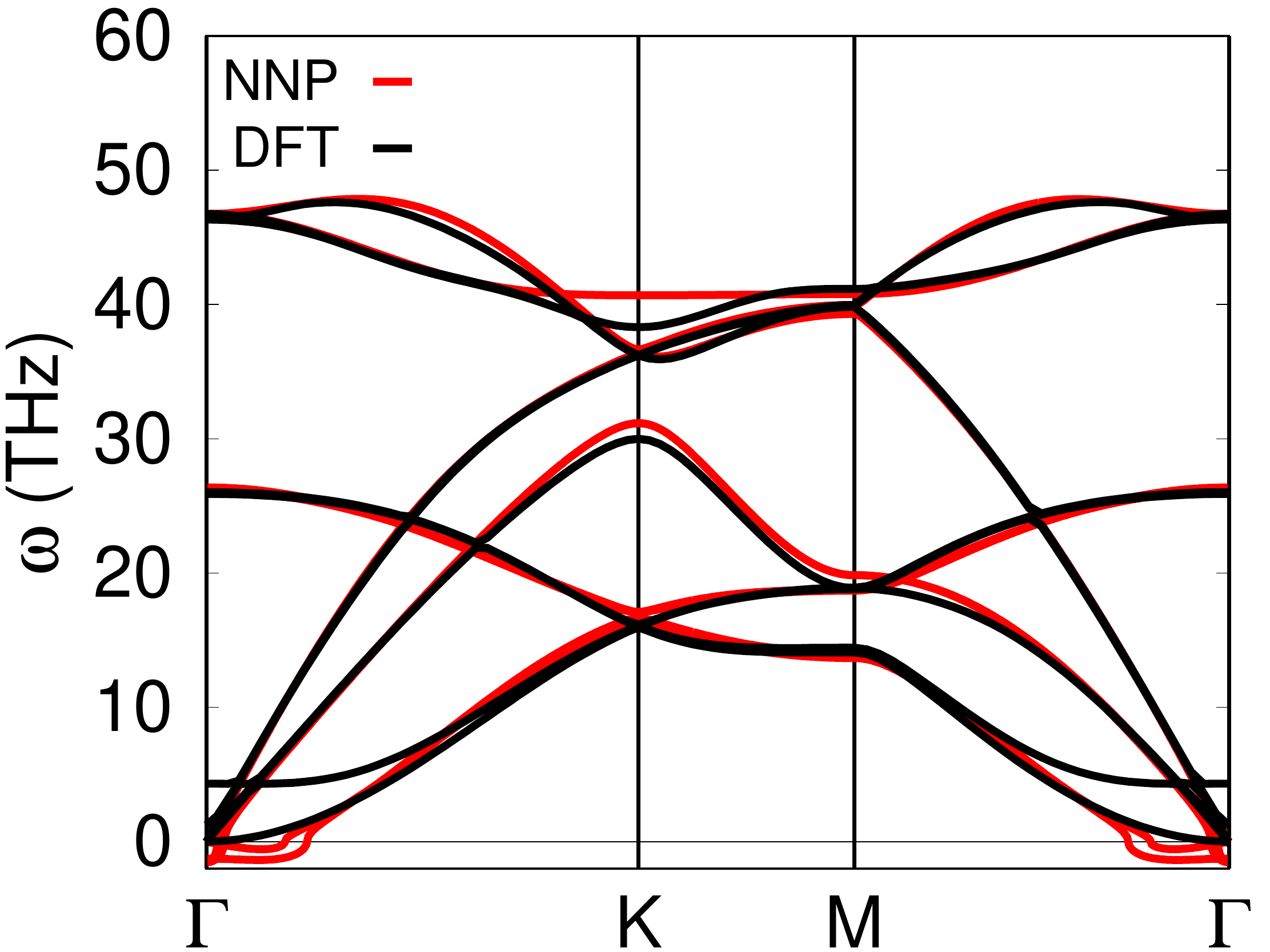}
        \includegraphics[width=0.14\textwidth]{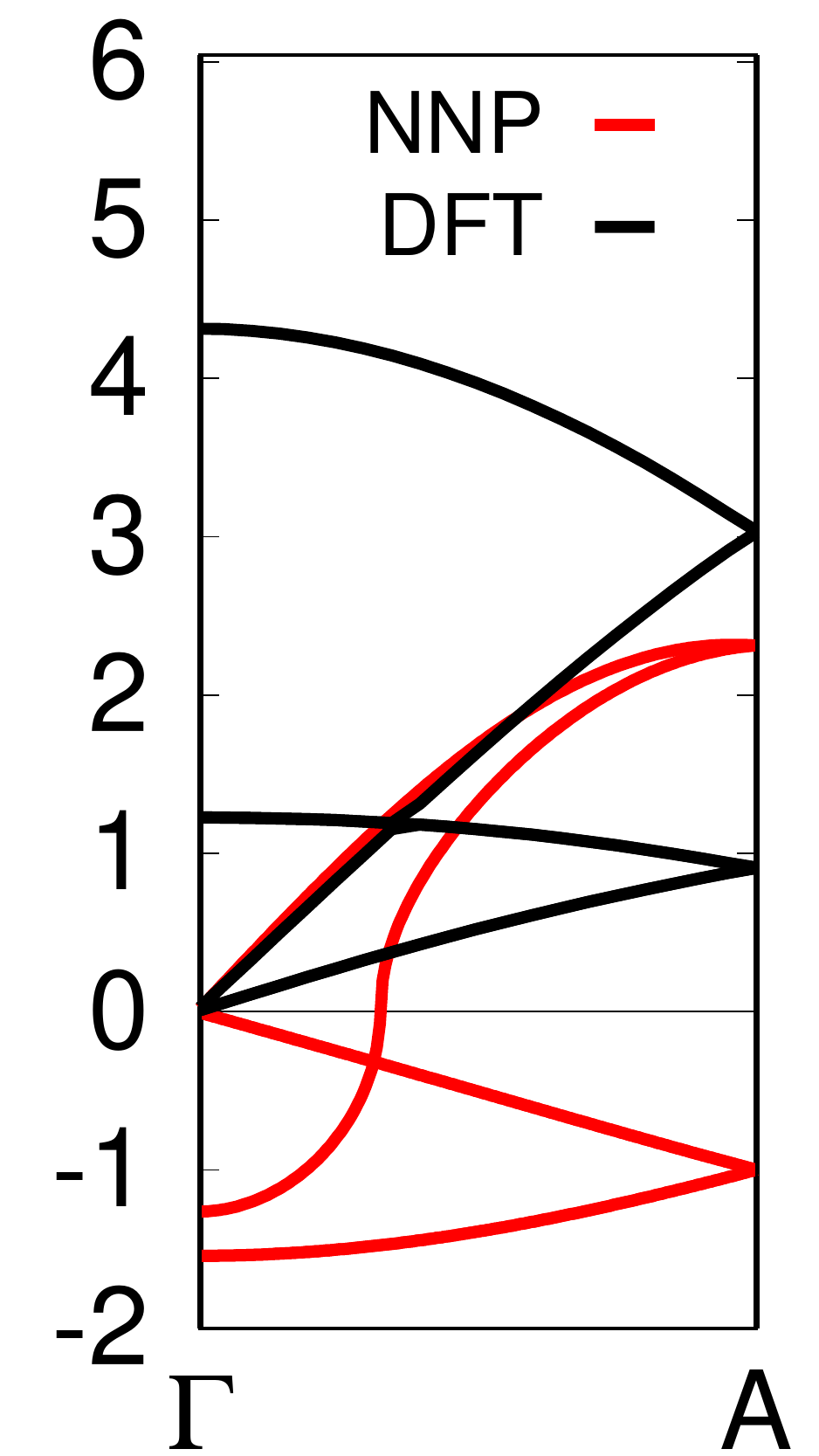}
    \par
        \includegraphics[width=0.33\textwidth]{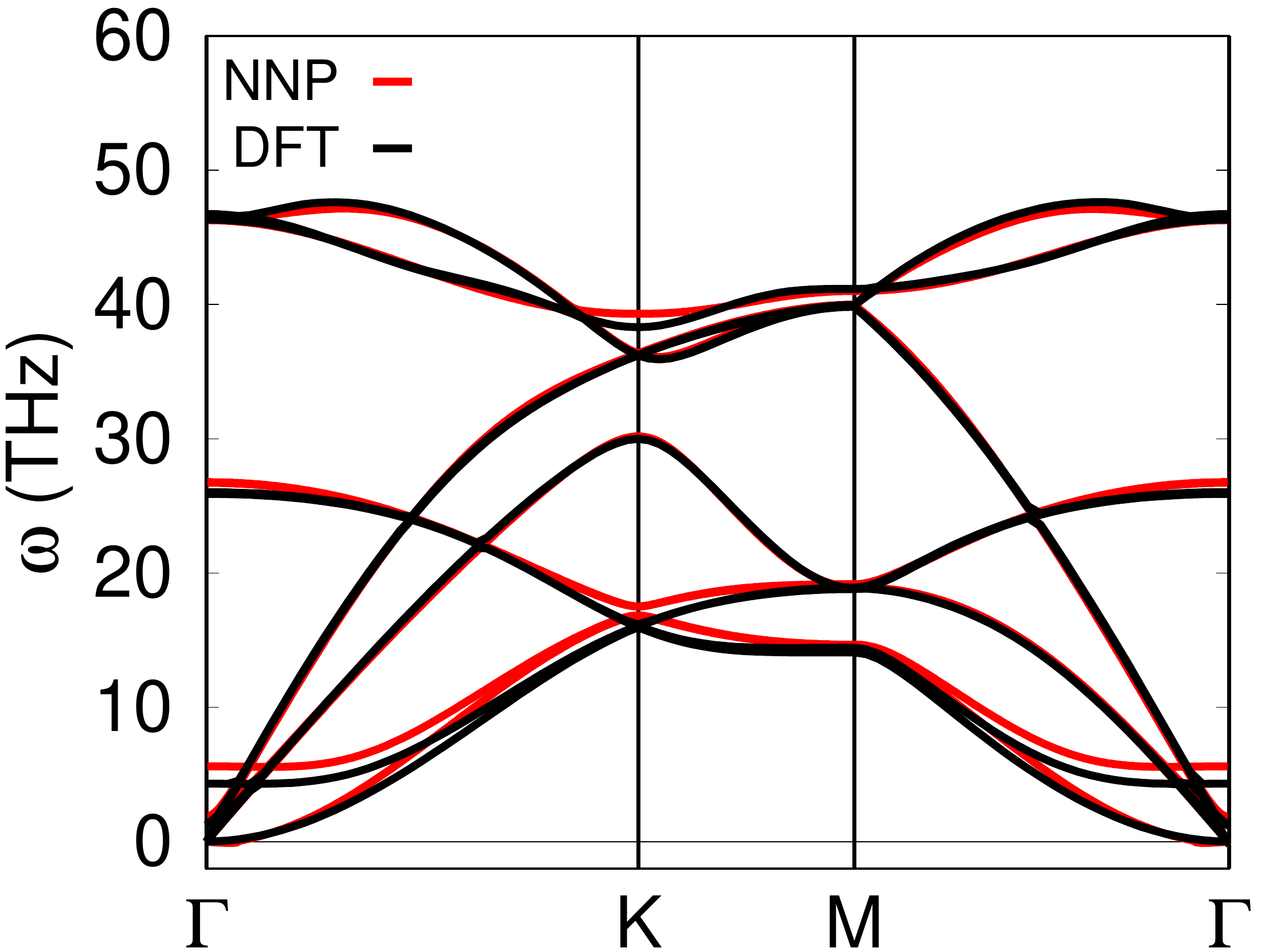}
        \includegraphics[width=0.14\textwidth]{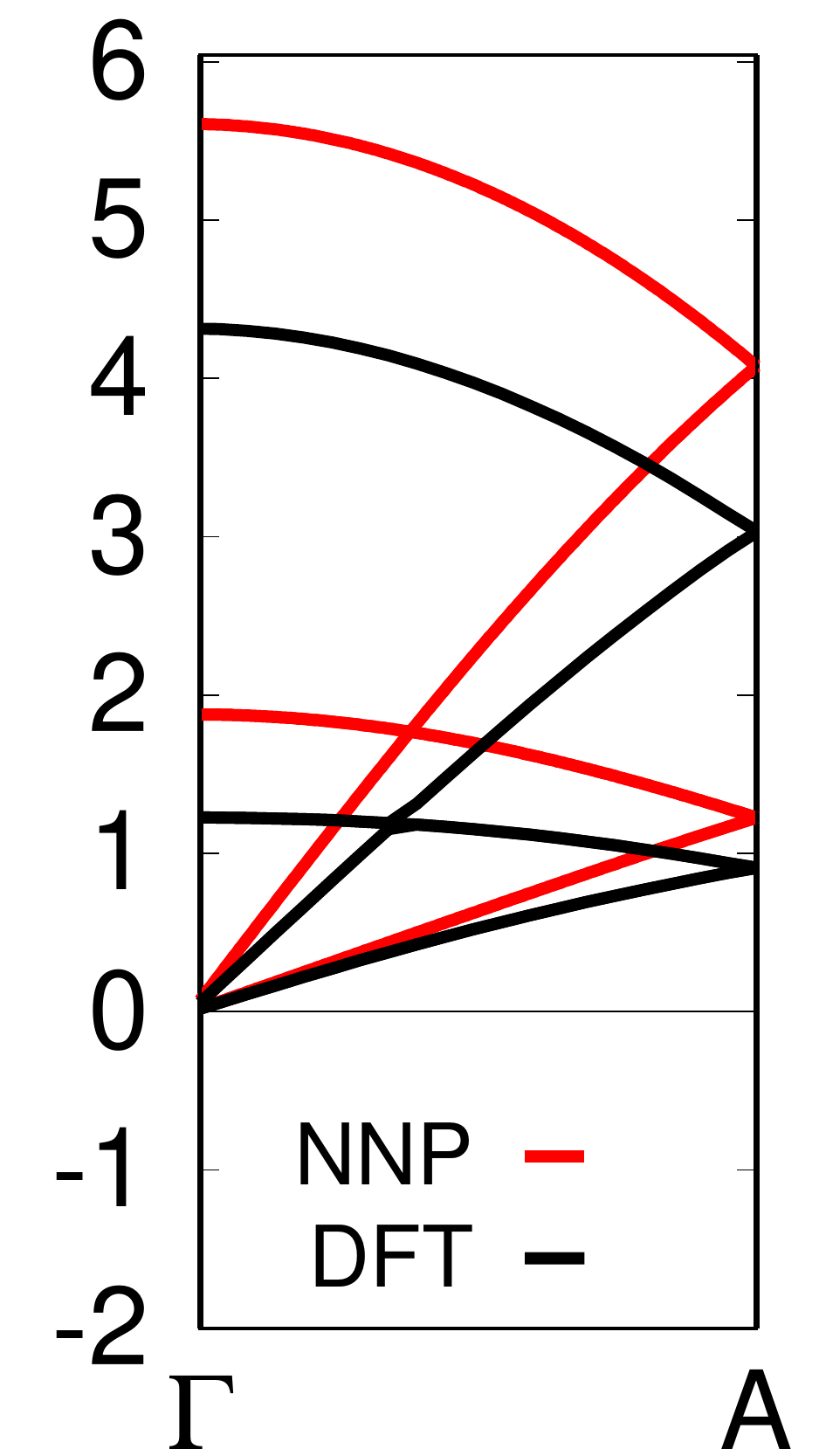}
    \caption{Phonon dispersion of graphite along the high symmetry lines for ({\bf top}) a NNP trained with the whole dataset at the last iteration and ({\bf bottom}) a NNP trained with all the data within $D=0.05$ from diamond and graphite ($D_{12}$, as described in Sec.~\ref{sec:EstimatingAccNNP}). The (small) imaginary frequencies are lifted by training on data close to graphite in structure. }
    \label{fig:phonon_graphite}
\end{figure}

Phonon dispersion relations give a complete picture of the elastic properties of a material, and reproduction of the dispersion relations obtained via DFT is a tight accuracy criterion on model potentials. 
Here we examine the performance of NNP through its prediction of phonon dispersion in the case of diamond and graphene, as a function of lattice parameter, up to a 1\% deviation from the equilibrium structure. 
This is a relevant range for thermal expansion of these materials as, for instance, the change in lattice parameter of diamond at temperatures up to 2000\,K is found to be below 1\,$\%$~\cite{JACOBSON2019}. 
Similarly, thermal expansion increases graphene lattice parameter only within 1\% at temperatures up to 2500\,K~\cite{Pozzo2011}.

The predictions of NNP for phonon dispersion of diamond and graphene are depicted in Fig.~\ref{fig:diamond_phonon}. 
There is an overall good agreement between NNP and DFT in the case of diamond. In the case of graphene, there is a slight disagreement for the transverse optical mode around K point. 
This is the same trend observed in other machine learnt potentials~\cite{RowePatrick2018,Wen2019} and likely the result of electronic structural properties associated with this special point coupling with the lattice vibration. 
For both structures, the predicted phonon frequencies reduce when the crystal expands and increase when it is compressed, as expected. An exception to this is the soft flexural mode of graphene close to $\Gamma$ point. 
The instability of graphene upon compression can be seen via small imaginary frequency of this mode (shown as negative). This feature is predicted with DFT and is successfully reproduced with NNP, pointing at the capacity of NNP in predicting important structural stability indicators.

Phonon dispersion of graphite, shown in Fig.~\ref{fig:phonon_graphite} displays negative frequencies for low wave vectors close to $\Gamma$, along the perpendicular direction to the graphene plane. 
These phonon modes are particularly soft and are very sensitive to the level of accuracy of the forces predicted by NNP. In order to verify this hypothesis, we retrain a NNP model (from scratch) using data from small neighborhoods of diamond or graphite (within a distance of 0.05 as defined in equation~\ref{structure_distances} of the Discussion section). There are no imaginary frequencies for this potential and a general agreement with DFT is obtained. 
However, as it will be further examined later (see Discussion), this NNP model only predicts properties of configurations around diamond and graphite and it is found to be highly non-transferable to other regions of the potential energy surface of carbon.   
\subsection{Amorphous carbon structures}
\begin{figure*}[hbt!]
    \centering
    \includegraphics[width=\textwidth]{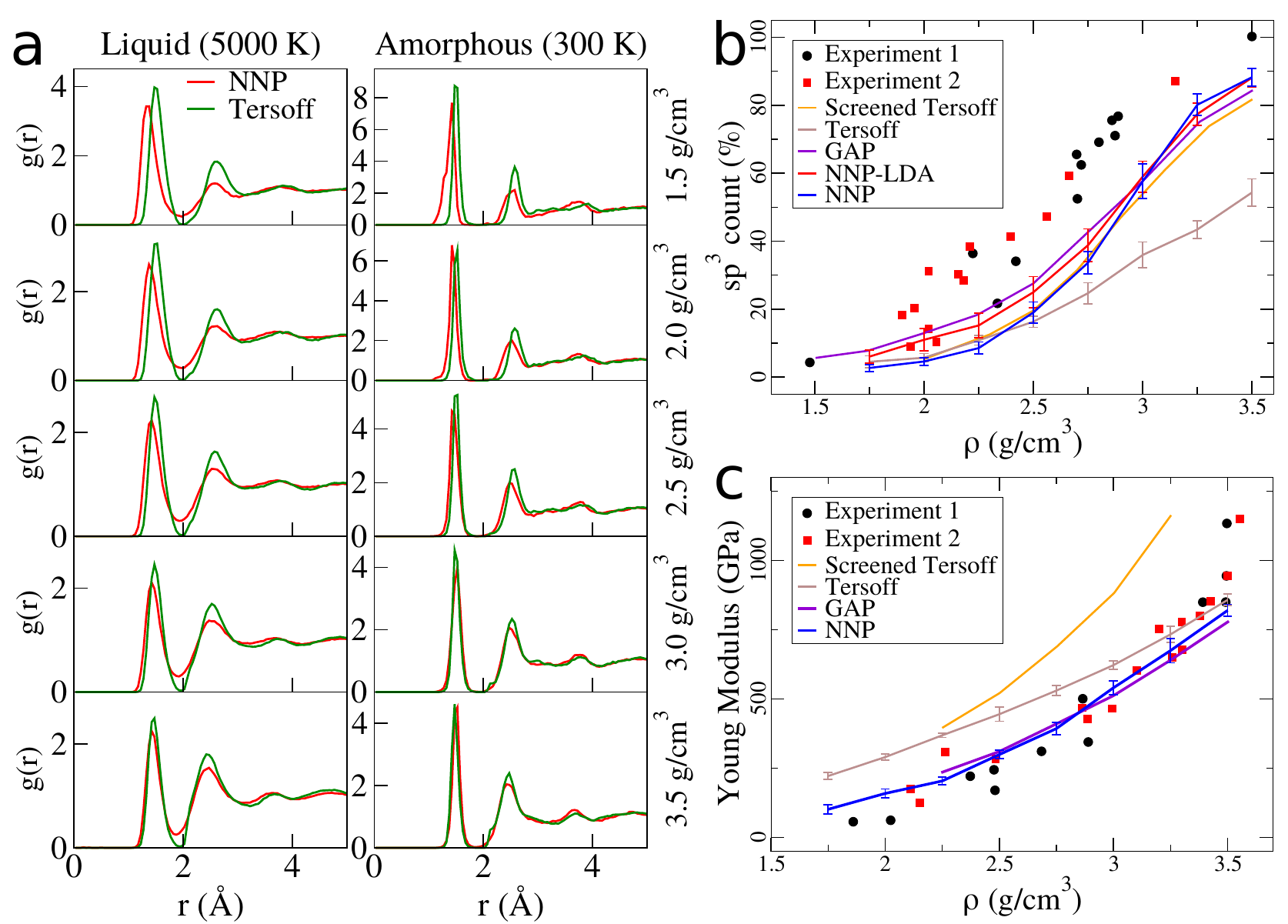}
    \caption{The amorphous phase of carbon:
   ({\bf a}) Radial distribution function for liquid (left) and amorphous (right) carbon, for our NNP and Tersoff potential, at increasing densities (top to bottom). ({\bf b}) Percentage of tetrahedrally coordinated atoms in amorphous carbon structures as a function of density, comparing NNP with rVV10 and LDA-level, and Tersoff potential to results taken from Ref.~\onlinecite{Deringer_amorphous_2017} for GAP and screened Tersoff potentials, as well as experimental results from Refs.~\onlinecite{Fallon_sp3} and \onlinecite{Ulrich_sp3}. ({\bf c}) Young Modulus of amorphous carbon as a function of density for  NNP at rVV10 level and Tersoff, compared to results taken from Ref.~\onlinecite{Deringer_amorphous_2017} for GAP and screened Tersoff potentials, as well as experimental results from Refs.~\onlinecite{SCHULTRICH1996} and \onlinecite{SCHULTRICH1998} }
    \label{fig:amorphous}
\end{figure*}


Lastly, we test the NNP in its ability to construct amorphous carbon structures in a range of densities from 1.5 to 3.5 g/cm$^3$ generated via the melt and quench method following the steps highlighted in Ref.~\onlinecite{Deringer_amorphous_2017}. 
We start from a 216 atoms simple-cubic simulation cell and randomized velocities at 9000\,K and perform molecular dynamics simulation first at 9000\,K with Nose-Hoover thermostat~\cite{NoseHoover} for 4 picoseconds (ps), followed by another at 5000\,K for 4\,ps, then a fast exponential quench to 300\,K at a rate of 10\,K/fs (total duration $\sim$0.5\,ps), and finally for 4\,ps we let the system evolve with the thermostat fixed at 300\,K. 

The radial distribution function (RDF) of liquid and amorphous phases are given in Fig.~\ref{fig:amorphous}{\bf a}. The liquid is less ordered than the amorphous configurations at all densities, for all potentials considered. 
In Ref.~\onlinecite{Deringer_amorphous_2017}, it was shown that both DFT and GAP have a non-zero first minimum for the liquid phase at about 1.9\,\AA{} which is not properly described by the screened Tersoff potential~\cite{scrtersoff}. 
Similarly, the NNP of this work captures the non-zero first minimum in the liquid phase. The density dependence of the RDF found in this work agrees well with the one reported in Ref.~\onlinecite{Deringer_amorphous_2017} as well.

In order to quantify the short-range order of amorphous structures, we calculate the $sp^3$ concentration by computing the fraction of carbon atoms with at least four neighbours within a 1.85\,\AA{} radius. 
In Fig.~\ref{fig:amorphous}{\bf b} we show the behavior of this quantity as a function of density, comparing with the results of Ref.~\onlinecite{Deringer_amorphous_2017} and those obtained with regular and screened Tersoff potentials~\cite{scrtersoff}. 
All methods underestimate the experimental observations yet show a similar general trend with density. 

There are quantitative differences among the predictions of theoretical models, in particular, the difference between NNP and GAP predictions are more significant at medium and low densities.
This may be attributed to the fact that the DFT dataset used to construct the GAP potential is built with local density approximation (LDA), while in this study the DFT dataset for NNP is built with an accurate exchange-correlation functional that includes van der Waals (vdW) interaction from first principles. 
In the low density region, vdW interactions allow bonding beyond the typical $sp^3$ bond length, such that low energy configurations can be constructed with less $sp^3$ and more $sp^2$ bonds; while at high densities and at  shorter length scales vdW interactions are of lesser significance. 
This is more evident as we compare the $sp^3$ count predicted with NNP-LDA as it agrees more closely with the GAP result, revealing the role of the underlying DFT reference in the prediction of the properties of amorphous materials with machine learnt potential models. 

The bonding character between atoms strongly affects the elastic properties of materials. Hence, comparing the elastic properties as observed by experiments with those predicted by theory is another way of assessing the theoretical prediction of $sp^3$ count in amorphous structures. 
In order to do that we first find the metastable configurations closest in the phase space to the amorphous structures examined so far, by further quenching the dynamics from 300\,K to 0\,K, and then performing geometry relaxation until the force components on atoms are below 1\,mRy/bohr at fixed volume. 
Fig.~\ref{fig:amorphous}{\bf c} shows the Young's modulus of these metastable amorphous structures as a function of density. 
The agreement with the experiment is remarkable, hinting that the discrepancy in theoretical and experimental $sp^3$ count seen in Fig.~\ref{fig:amorphous}{\bf b} might stem from an inconsistency in definitions between theory and experiment, i.e. the neighbor count within 1.85\,\AA{} used in theory underestimates the experimentally measured value that is obtained via comparison of electron energy-loss spectroscopy (EELS) peak area to graphitized carbon~\cite{Fallon_sp3,Ulrich_sp3}.

\section{Discussion}
\label{sec:EstimatingAccNNP}

The accuracy of a neural network model is often measured by the distribution of the prediction error on a test dataset, in particular via mean and standard deviation of error. 
But as is the case with training sets, test sets are also not standardized between studies. Therefore the accuracy of potentials tested on different datasets cannot be compared. 
Here we study the effect of the training and test sets on the apparent accuracy of networks, and measure the impact of these sets on the transferability of neural network potentials.

For every configuration in a dataset we first define its Euclidean distance from a reference atomic environment (e.g.~cubic diamond, graphite). 
The distance between the reference configuration $\alpha$ and a given configuration $\beta$ is defined as 
\begin{equation} \label{structure_distances}
d_{\alpha \beta}=\frac{1}{2}\left(\frac{1}{N^{\rm at}_\beta} \sum_{i=1}^{N^{\rm at}_\beta} |{\bf g}_\alpha - {\bf g}^i_\beta|^2\right)^{1/2}
\end{equation}
where  ${\bf g=\frac{G}{|G|}}$ with ${\bf G}$ being a ``fingerprint" vector that describes the atomic environment of all atoms in the unit cell for a given configuration, $N^{\rm at}_\beta$ is the number of atoms in configuration $\beta$. 
In this work, for the definition of atomic environment, we use the well-established atom-centered symmetry functions of Behler and Parrinello \cite{Behler2007}, with modifications by Ref.~\onlinecite{ANI17} and \onlinecite{lot2019panna}. This definition is also used to describe the input to the neural network architecture. (See Methods section for a detail description of the description vectors and their use in neural network training.)

Then, we construct a dataset by considering only configurations within a given cutoff distance $D$ from this reference. 
Following this strategy we build four datasets, three of which are referenced from cubic diamond with $D$ values of 0.05, 0.10 and 0.15; the fourth one is referenced from either cubic diamond or graphite with $D$=0.05 (denoted by $D_{12}$). 
For each D, 20\,$\%$ of the dataset is set aside for validation and the remaining 80\,$\%$ is used for training. We train four different NNPs on these four sets from scratch, and test each on the respective validation datasets.

\begin{figure}[!hbt]
    \centering
    \includegraphics[width=0.48\textwidth]{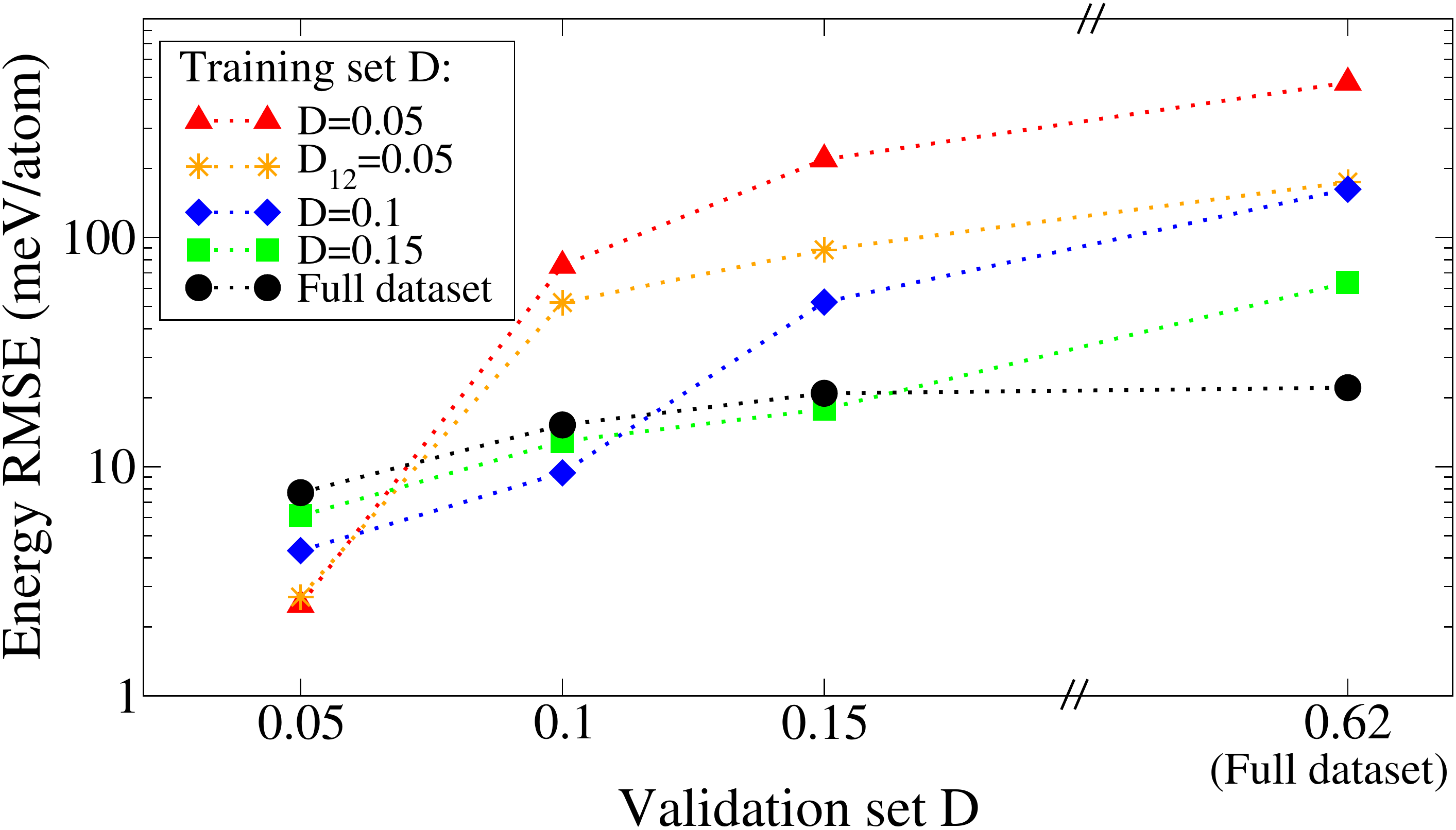}\\[10pt]
    \includegraphics[width=0.48\textwidth]{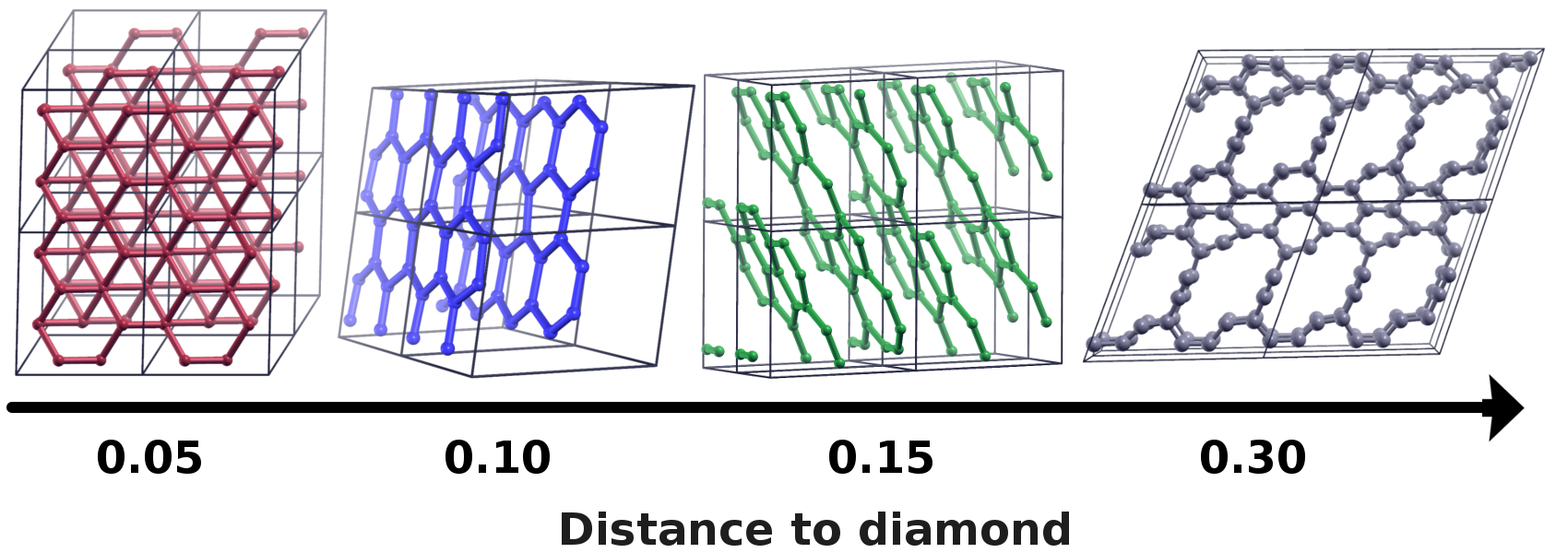}\\[5pt]
    \includegraphics[width=0.48\textwidth]{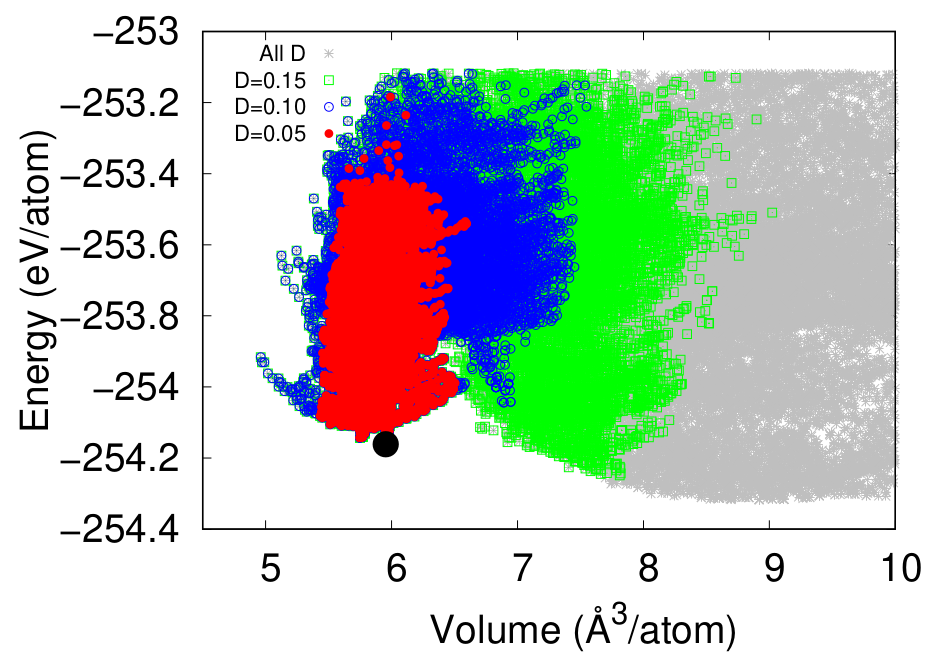}
    \caption{{\bf Top:} Validation error of networks trained on different datasets as a function of the distance of the validation set from diamond. Numerical values are given in the Supplementary Material for energies and forces (not shown here). 
    {\bf Middle:} Representative structures at given distances from diamond, the reference structure. The structures at 0.05 or lower are recognizably related to the reference, while at 0.10 and 0.15 compressed and/or defected layered structures are visible. At 0.30 and above, configurations with several double bonds and Carbon chains appear. 
    {\bf Bottom:} Energy per atom as a function of volume for structures in the dataset, colored according to their distance cutoff $D$ from diamond. The black dot corresponds to the reference diamond structure. The complete dataset includes structures with larger volume that are omitted here for clarity. The complete volume range is given in the Supplementary Material.}
    \label{fig:valdist}
\end{figure}

In the top panel of Fig.\ref{fig:valdist}, we report the training and validation RMSE in energy prediction as the cutoff distance $D$ from the reference structure increases. 
We show that an RMSE as low as 2.4 (2.5) meV/atom for training (validation) can be obtained when training and validation configurations are very similar, i.e.\ within a distance of 0.05 from the diamond reference. 
However, the prediction error of this NNP dramatically increases as it gets tested on structures farther in the input space, to as high as an RMSE of 473 meV/atom. This is a confirmation of the common observation that the prediction error of a neural network is strongly dependent on the similarity of training and test environments~\cite{Bernstein2020}. 
On the other hand, when the model is trained and tested using the complete set, a prediction RMSE of 22.1\,meV/atom is obtained for energy,  while, for the configurations within $D=0.05$ from diamond, the prediction RMSE is still considerably small, 7.7\,meV/atom.  The analysis for forces follows the same trend as energies. The RMSE values for energies and forces are given in the Supplementary Material.

Hence, it can be deduced that, for a fixed  network architecture, a trade-off must be struck between having small error on configurations similar to a reference structure, and obtaining reliable predictions for general configurations from the full potential energy surface. The other entries in these tables confirm this analysis: the more diverse the training set is, the more robust is the resulting potential outside its training basin. Therefore, for a reliable NNP for multiple C polymorphs, as the one targeted here, a diverse training set from a wide region of the potential energy surface is necessary. 

In summary, in this work we have presented a self-consistent technique for generating an accurate and transferable neural network potential. 
Since neural networks encode the physics of a system into their parametrization through data, the dataset plays a crucial role of the resulting NNP performance.
The self consistent method described in this work achieves an accurate and comprehensive dataset via balanced integration of evolutionary algorithm, unsupervised machine learning in the form of clustering, and molecular dynamics.  

The presented workflow for the self consistent refinement of the potential requires minimum human intervention. 
Therefore not only it is ready for high-throughput automation schemes as envisioned in future of experimentation but it is also robust with respect to lack of previous information about a system, as is often the case with novel materials.
The workflow and the underlying neural network~\cite{panna_code} and electronic structure codes are publicly available and are open-source.

The self-consistent NNP generation procedure is entirely system independent and we demonstrated its successful application to the challenging case of Carbon for which classical and machine-learnt potentials are abundant in literature. 
We show that for diamond, graphite and graphene phases, NNP reported in this work performs considerably better than Tersoff, a classical potential, and overall better than the existing machine learnt potentials for structural and elastic properties. 
When predicting graphite phonon dispersion, NNP resulted in very good agreement for the majority of the modes, yet predicted instability for the  very soft modes that relate to interlayer interaction. 
We have traced this behaviour to the accuracy requirement in predicting such small forces. To increase accuracy using a fixed neural network architecture, we built the training set only with structures that are in the vicinity of graphite according to a fingerprint based distance measure. 
The resulting potential provided accurate phonon frequencies but it showed poor generalization to a wider range of structures, compared to a more comprehensive potential trained on the entire dataset. 
This example highlights the need for a procedure to standardize the accuracy measure of NNPs and a more pressing need to build error estimate measures into the process of generating NNPs. 

\section{Methods}

\subsection{Evolutionary Algorithm for Configuration Space Search}
We start the self consistent cycle using ReaxFF~\cite{Raju2015}, a reactive FF, that was fitted for Li-C system, to generate the initial configurations with 16 and 24 Carbon atoms per unit cell at 0, 10, 20, 30, 40 and 50 GPa via EA as implemented in USPEX~\cite{USPEX2006,Oganov2006}. 
At each pressure, we start with a population of 30 (50) randomly generated structures for the 16 (24) atoms per unit cell,
and evolve it through the following evolutionary operations with the given ratios: heredity (two parent structures are combined) 50$\%$, mutation (a distortion matrix is applied to a structure) 25$\%$, or by generating new random structures 25$\%$. 

At each generation, structures are optimized in five successive steps: 
(a) constant pressure and temperature molecular dynamics at 0.1\,GPa and 50\,K respectively for 0.3\,ps  with time step of 0.1\,fs, 
(b) relaxation of cell parameters and internal coordinates until force components are less than 0.26\,eV/\AA{}, 
(c) constant pressure and temperature molecular dynamics at 0.1\,GPa and 50\,K respectively for 0.3\,ps  with time step of 0.1\,fs, 
(d) relaxation of cell parameters and internal coordinates until force components are less than 0.026\,eV/\AA{},  and 
(e) a final relaxation of cell parameters and internal coordinates until force components are less than 0.0026\,eV/\AA{}. 

Only the 70$\%$ most energetically stable ``parents" were allowed to participate in the process of creating the new generation. 
In the heredity step, only sufficiently distinct structures (whose cosine distance, as defined in the next section, is greater than a given threshold) are considered as parents. 
This threshold is fixed at 0.008 in the first iteration, as it is small enough to allow deformed structures from the same polymorph to be parents. 
In order to enhance the diversity of the structures in the subsequent iterations, the threshold is increased to 0.05 so that the parents can be expected to be from different polymorphs. 

Each structure search is evolved up to a maximum of 50 generations at the first iterations and 30 in the subsequent ones. 
The configuration space search performed this way produces a wide range of $sp^2$, $sp^3$ and mixture of $sp^2$ and $sp^3$ structures, including defective layered structures. 

\subsection{Clustering}
Initially, an unsupervised, bottom-up,  distance-based hierarchical clustering approach with single linkage 
is used on all structures obtained with EA to identify the unique polymorphs. 
In the later iterations, clustering is applied only to those structures where NNP prediction differs from DFT ground truth energy by more than 5~meV/atom. 
That way, polymorphs that are already well described by NNP are not over-sampled.
During clustering, to measure the similarity between structures, we use the fingerprint-based cosine distance defined in Ref.s~\onlinecite{Oganov2009} and~\onlinecite{Valle2010}. 
In the case of a single species in the unit cell, and in its discretized form, the fingerprint of a configuration becomes: 
\begin{equation}
     F[k] = \frac{1}{2}\sum_{i, {\rm cell}} \sum_{j} \frac{{\rm erf} \left[ \frac{(k+1)\Delta-R_{ij}}{\sqrt{2}\sigma}\right]-{\rm erf}\left[\frac{k\Delta-R_{ij}}{\sqrt{2}\sigma}\right]}{4\pi R_{ij}^2 \frac{N^2 }{V}\Delta} -1
 \end{equation}
where the first sum runs over all atoms $i$ in the unit cell and the second sum runs over all atoms $j$ within a spherical cutoff radius $R_{\rm max}$, and  $R_{ij}$ is the distance between atoms $i$ and $j$.  
The numerator describes the integral of a Gaussian density of width sigma over a bin of size $\Delta$. $N$ is the number of atoms in the unit cell and $V$ is the unit cell volume.

The cosine distance between structures 1 and 2 is defined as
\begin{equation}
    D_{\rm cosine}(1,2) = \frac{1}{2}\left(1-\frac{{\bf F_1}\cdot
    {\bf F_2} }{|{\bf F_1}||{\bf F_2}|}\right) .
\end{equation}
The dimension of the $F$-vector is set to $R_{\rm max}/\Delta=125$ with $R_{\rm max}=10$\AA{} and $\Delta=0.08$ in this work.
Two configurations closer to one another than a distance threshold are determined to belong to the same cluster. 
In this work the threshold is tuned to yield approximately 100-150 clusters at each step, which results in affordable computational cost for the remaining calculations of the self-consistent cycle.

\subsection{Molecular Dynamics}
We manually select a representative structure from each cluster and perform a 0.5~ns classical NPT molecular dynamics simulation with Nose-Hoover thermostat and barostat. 
In these simulations the external conditions of pressure and temperature are ramped up from -50GPa at 100 K, to 50GPa at 1000K in the course of 0.5~ns. 
The characteristic relaxation times of the thermostat and barostat are chosen as 50\,fs and 100\,fs, respectively.
By sampling a snapshot of the dynamics every 5 ps, 100 configurations are selected. All molecular dynamics simulations are performed with LAMMPS package~\cite{LAMMPS19951}.
In addition, 440 randomly selected graphene atomic configurations from the libAtoms repository~\cite{WebHomeHomelibAtomsorg} are added to the selection. 
This set constitutes the set of structures where {\it ab initio} total energy calculations are then performed and added to the training set.

\subsection{First Principles Calculations}
The first principles calculations performed on all the structures visited during EA configuration space search and MD refinement described earlier employ the following parameters: 
Plane wave basis set kinetic energy cutoff for wavefunctions and charge density is 80 and 480 Ry respectively. 
The  rVV10~\cite{rvv10} exchange-correlation functional that incorporates non-local van der Waals correlations is employed. 
A Brillouin zone sampling with resolution of 0.034$\times 2 \pi$ \AA{}$^{-1}$ for the 3D carbon structures and 0.014$\times 2 \pi$ \AA{}$^{-1}$ for graphene is used. 
These parameters are found to yield 1mRy/atom precision on diamond, graphite and graphene. All DFT calculations were performed with the Quantum ESPRESSO package~\cite{QE-2009,QE-2017}. 
Elastic properties are computed through the thermopw framework~\cite{thermo_pw} while vibrational properties are obtained with PHON package~\cite{ALFE2009}.

In the first self-consistent iteration, the training set is made up of all generated structures lying within 10\,eV from the lowest energy one. 
This results in a total of $\sim$16000 configurations. 
In the subsequent iterations of the self-consistent procedure, we use all configurations whose energy per atom is within 1.2\,eV of the lowest one,
these are added to the previously selected configurations, amounting to a total of about 30000 configurations in the second and 60000 configurations in the third and final iteration. From these configurations, 20\,\% was set aside for validation and the remaining 80\,\% was used in the NNP training. 

\subsection{Neural Network Architecture}
In this work we adopt the Behler-Parrinello approach to atomistic neural networks~\cite{Behler2007} where the total energy of a system of $N$ atoms is defined as the sum of atomic energy contributions:
\begin{equation}\label{energy_eq}
    E = \sum_{i=1}^N E_{i}(G_i),
\end{equation}
where $E_{i}$ is the energy contribution of an atom $i$, and $G_i$ is its local environment descriptor vector. 
As described in detail in the next section, we choose descriptors with $144$ components per atomic environment. 
The contribution of an atom to the total energy is obtained by feeding its environment descriptor to the feed-forward all-to-all-connected neural network. 
Here we build a network with two hidden layers, with 64 and 32 nodes for the first and second layer respectively, both with Gaussian activation function, and a single-node output layer with linear activation. 
The resulting network has a total of $ 11393$ parameters, i.e. $(144\times64)+(64\times32)+(32\times1)=11296$ weights and $64+32+1=97$ biases. 
The energy of each atom is then summed to obtain the total energy of the configuration.  
The force on each atom can be obtained analytically:
\begin{equation}\label{force_eq}
    {\bf F}_i = - \sum_{j}\sum_\mu \frac{\partial E_{j}}{\partial G_{j\mu}} \frac{\partial G_{j\mu}}{\partial {\bf R}_{i} }
\end{equation}
where the atom index, $j$, runs over all the atoms within the cutoff distance of atom $i$, and index $\mu$ runs over the descriptor components. 

During training, the weight and bias parameters $W$, are optimized with the Adam algorithm~\cite{Adam} using gradients obtained by randomly selected subsets (minibatches) of data. 
The loss function of this stochastic optimization problem is defined as the sum of two contributions: one using the total energy value (Eq.~\ref{eq:Eloss}) and one using the force on each atom (Eq.~\ref{eq:Floss}):
\begin{align}\label{eq:Eloss}
    {\cal L}^{\rm E}(W) & = \sum_{c\in {\rm batch}} \left(E^{\rm DFT}_c-E_c(W) \right)^2\\
               & +  \exp\left[ a \tanh \left(\frac{1}{a} \sum_{c\in {\rm batch}}\left(\frac{E^{\rm DFT}_c-E_c(W)}{N_c} \right)^2\right)\right ]\nonumber ,
\end{align}
where $E_c^{\rm DFT}$ is the ground-truth total energy obtained via DFT and $E_c$ is the NN prediction for total energy of a given configuration $c$, consisting of $N_c$ atoms in the unit cell. 
The second part of this equation exponentially penalizes outliers while keeping the exponent normalized; $a$ is a constant that allows to tune this penalty, $a=5$ is used in this study.
The force contribution to the loss is given by:
\begin{equation}\label{eq:Floss}
    {\cal L}^F(W) = \gamma_F \sum_{c \in \rm{batch}}\sum_{i=1}^{N_c} 
     \left|{\bf F}^{\rm DFT}_{i} - {\bf F}_{i}\right|^2,
\end{equation}
where for any atom $i$ of configuration $c$, ${\bf F}^{\rm DFT}_{i}$ is the ground-truth force obtained via DFT, and ${\bf F}_{i}$ is the NN prediction for it. $\gamma_F$ is a user-defined parameter that controls the scale of this loss component. 
The results reported are obtained with $\gamma_F$ equals 0.5. 

An $L_2$-norm regularization term is also added with a small coefficient $\gamma_R=10^{-4}$ to prevent weights from becoming spuriously large: 
\begin{equation}
    {\cal L}^R(W) = \gamma_R \frac{|W|^2}{2}.
\end{equation}

The total loss is thus defined as
\begin{equation}
    {\cal L} (W) = {\cal L}^{\rm E}(W) +  {\cal L}^F(W) + {\cal L}^R(W).
\end{equation}

All models are trained starting from random weights and a starting learning rate $\alpha_0 = 0.001$. 
The learning rate is decreased exponentially with optimization step $t$ following the relationship $\alpha(t) = \alpha_0 r^{t/\tau}$ with decay rate $r=0.96 $ and the decay step $\tau=3200$. 
A batch size of 128 data points is used throughout the study.

\subsection{Atomic Environment Descriptors}
We use Behler-Parrinello symmetry functions~\cite{Behler2007} as local atomic descriptors. These functions include a two body and a three-body term, referred to as radial and angular descriptor, respectively. 
We use a modified version of the original angular descriptor~\cite{ANI17} as implemented and detailed in PANNA package~\cite{lot2019panna}. The radial descriptor function is defined as:
\begin{equation}\label{eq:BPrad}
    G_i^{\rm Rad}[s]=\sum_{j\neq i}
    e^{-\eta\left(R_{ij}-R_s\right)^2}
    f_c(R_{ij}),
\end{equation}
where $\eta$ and a set of Gaussian-centers $R_s$ are user-defined parameters of the descriptor. The sum over $j$ runs over all atoms whose distance $R_{ij}$ from the central atom $i$ is within the cutoff distance $R_c$. The cutoff function, $f_c$ is defined as:
\begin{equation}\label{eq:fc}
    f_c(R_{ij})=\left\{
    \begin{array}{ll}
         \frac{1}{2}\left[
            \cos\left(\frac{\pi R_{ij}}{R_c}\right)+1
         \right]&R_{ij}\leq R_c  \\
         0& R_{ij}>R_c.
    \end{array}
    \right.
\end{equation}
The angular part of the descriptor with central atom $i$ is defined as:
\begin{align}
\label{eq:BPang}
G_i^{\rm Ang}[s]=&2^{1-\zeta}\sum_{j,k\neq i}
    \left(1+ \cos (\theta_{ijk} - \theta_s) \right)^{\zeta} \nonumber \\
&\times e^{-\eta \left(R_{ij}/2+R_{ik}/2 -R_s\right)^2}\nonumber \\
&\times f_c(R_{ij})f_c(R_{ik}). 
\end{align}
The sum runs over all pairs of neighbours of atom $i$, indexed as $j$ and $k$, with distances $R_{ij}$ and $R_{ik}$ within the cutoff radius $R_c$, forming an angle $\theta_{ijk}$ with it. Here $\eta$, $\zeta$, and the sets of $\theta_s$ and $R_s$ are the user-defined parameters of the descriptor. 

We note that the descriptor as written in Eq.~\ref{eq:BPang} has discontinuous derivative with respect to atomic positions when atoms are collinear. 
To restore the continuity, we replace the $\cos(\theta_{ijk}-\theta_s)$  term with the following expression
\begin{equation}
2\frac
{\cos(\theta_{ijk})\cos(\theta_s)+\sqrt{1-\cos(\theta_{ijk})^2+\epsilon\sin(\theta_s)^2}\sin(\theta_s)}
{1+\sqrt{1+\epsilon\sin(\theta_s)^2}}    
\label{eq:epsilon}
\end{equation}
where we introduce a small normalization parameter, $\epsilon$, such that the expression approaches $\cos(\theta_{ijk}-\theta_s)$ in the limit of $\epsilon \rightarrow 0$. In this work, $\epsilon=0.001$ was used, while values between $0.001-0.01$ were found to yield stable dynamics and equivalent network potentials for any practical purpose. 

The radial descriptors are parametrized with $\eta = 16.0$ ~\AA$^{-2}$, while 32 equidistant Gaussian centers, $R_s$, are distributed between 0.5~\AA\ and 4.6~\AA. 
For the angular part $\eta=10.0$~\AA$^{-2}$, $\zeta=23.0$, 8 equidistant $R_s$ are distributed between 0.5~\AA\ and 4.0~\AA\, and 14 $\theta_s$ are chosen between $\pi/28$ and $27\pi/28$ with spacing $\pi/14$. 
The cutoff $R_c$ is 4.6~\AA\ for radial and 4.0~\AA\ for the angular descriptors, respectively. 
The resulting descriptor has a total of $32+14 \times 8 = 144$ components per atomic environment. 

%

%

\section{Data Availability}
The neural network potential described in this work is released in PANNA~\cite{lot2019panna} format compatible with several molecular dynamics packages via OPENKIM~\cite{openkim}. A native LAMMPS plugin version is also given in the Supplementary Material.

\section{Acknowledgements} 
 The work of E.\ Kaxiras and E.\ Kucukbenli was supported by a DOE grant, BES Award DE-SC0019300. E.\ Kucukbenli, F.\ Pellegrini and S.\ de Gironcoli are grateful for the financial support by European Union’s Horizon 2020 research and innovation program under grant agreement No.\ 676531 (project E-CAM). S.\ de Gironcoli also acknowledge EU funding under grant agreement No.\ 824143 (project MaX).  This work used the high performance computing resources of CINECA, SISSA and FASRC Cannon cluster supported by the FAS Division of Science Research Computing Group at Harvard University. This work  also used the Extreme Science and Engineering Discovery Environment (XSEDE), which is supported by National Science Foundation grant number ACI-1548562~\cite{xsede}, specifically it used Stampede2 at TACC through allocation TG-DMR120073.
 
\section{Author Contributions}
E.\ Kucukbenli and S.\ de Gironcoli designed and planned the study. E.\ Kucukbenli, F.\ Pellegrini, S.\ de Gironcoli supervised all aspects of the
project. R.\ Lot, F.\ Pellegrini, Y.\ Shaidu, E.\ Kucukbenli implemented the methodology into PANNA code, and performed
extensive tests. Y.\ Shaidu performed DFT calculations and constructed the ANN
potentials. Y.\ Shaidu, F.\ Pellegrini, S.\ de Gironcoli and E.\ Kucukbenli analyzed the results. Y.\ Shaidu, E.\ Kucukbenli and S.\ de Gironcoli led the manuscript writing. All authors contributed to discussions throughout the study and commented on the manuscript.

\section{Competing Interests}
The authors declare no competing interests.

\section{Additional Information}
Supplementary information is available. 
Correspondence and requests for materials should be addressed to E.\ Kucukbenli and S.\ de Gironcoli.
\newpage

\appendix
\bibliography{references}
\end{document}


\title{Supplementary Material for ``A Systematic Approach to Generating Accurate Neural Network Potentials: the Case of Carbon"}

\author{Yusuf Shaidu}
\affiliation{Scuola Internazionale Superiore di Studi Avanzati, Trieste, Italy.}
\affiliation{The Abdus Salam International Centre for Theoretical Physics, Trieste, Italy.}

\author{Emine K\"{u}\c{c}\"{u}kbenli}
\affiliation{John A. Paulson School of Engineering and Applied Sciences, Harvard   University, Cambridge, Massachusetts 02138, USA}
\affiliation{Scuola Internazionale Superiore di Studi Avanzati, Trieste, Italy.}

\author{Ruggero Lot}
\affiliation{Scuola Internazionale Superiore di Studi Avanzati, Trieste, Italy.}

\author{Franco Pellegrini}
\affiliation{Laboratoire de Physique de l'\'{E}cole normale sup\'{e}rieure, ENS, Universit\'{e} PSL, CNRS, Sorbonne Universit\'{e}, Universit\'{e} de Paris, F-75005 Paris, France}

\author{Efthimios Kaxiras}
\affiliation{John A. Paulson School of Engineering and Applied Sciences, Harvard University, Cambridge, Massachusetts 02138, USA}
\affiliation{Department of Physics, Harvard University, Cambridge, MA 02138, USA}

\author{Stefano  de Gironcoli}
\affiliation{Scuola Internazionale Superiore di Studi Avanzati, Trieste, Italy.}
\affiliation{CNR-IOM DEMOCRITOS, Istituto Officina dei Materiali, Trieste, Italy.}

\date{\today}

\maketitle

\section{Evolution of Training and Validation Error}

\begin{figure}[htb!]
\includegraphics[width=0.32\textwidth]{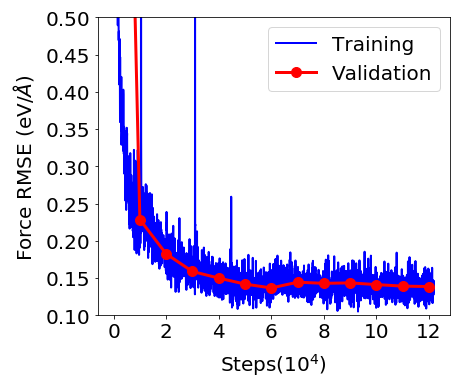}
\includegraphics[width=0.33\textwidth]{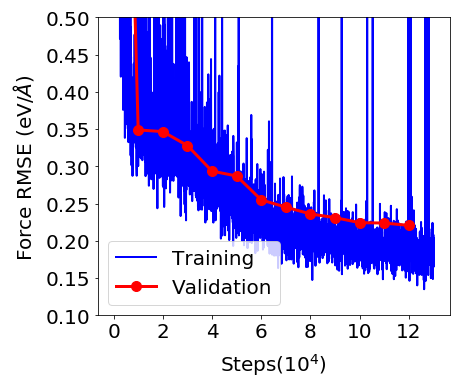}
\includegraphics[width=0.33\textwidth]{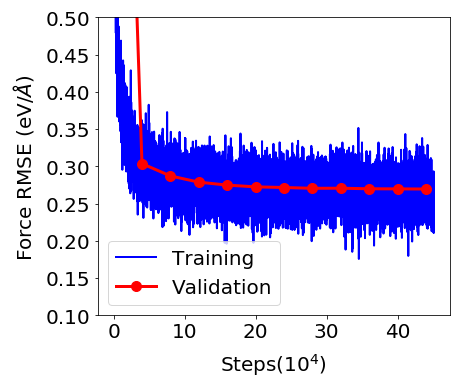}
\caption{ The evolution of error in forces on training and validation set for potentials trained at first (\textit{left}), second (\textit{middle}) and third iteration (\textit{right}) of the self-consistent cycle. The blue lines are the RMSE on a given batch of 128 configurations during training. The networks are evaluated during training on all the validation set of sizes $\approx$3000, $\approx$5200 and $\approx$12000 configurations for first, second and third iterations respectively (red dots with lines as guide to eye). }
\end{figure}

\begin{figure}[htb!]
        \includegraphics[width=0.32\textwidth]{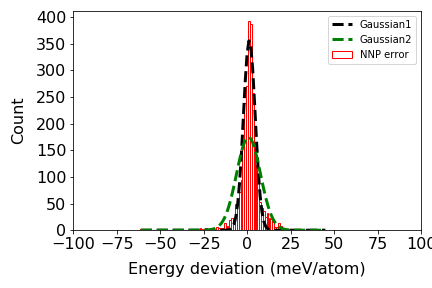}
        \includegraphics[width=0.33\textwidth]{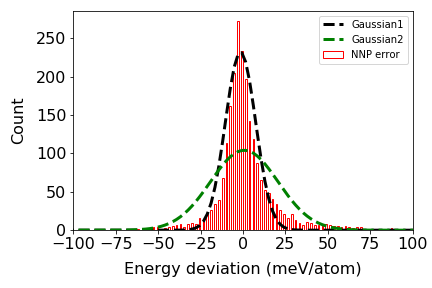}
        \includegraphics[width=0.33\textwidth]{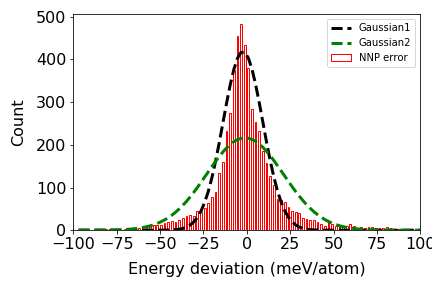}
          \includegraphics[width=0.32\textwidth]{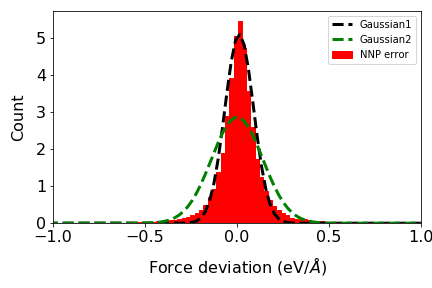}
        \includegraphics[width=0.33\textwidth]{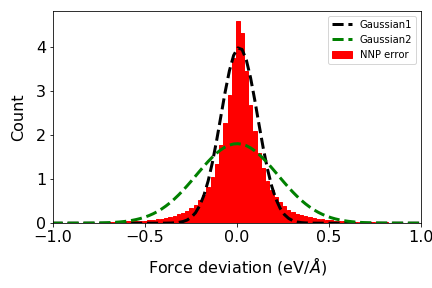}
        \includegraphics[width=0.33\textwidth]{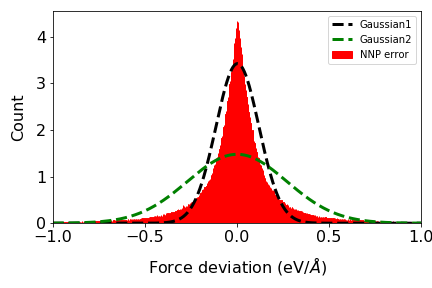}
   \caption{Error distribution for the validation dataset at first, second and third iteration from left to right for energies (top) and forces (bottom). The histogram is obtained with 3000, 5200 and 12000 configurations respectively. ({\bf Top}): The histogram is constructed such the area is equal to the number of configurations. The black dashed line is a normalized Gaussian fit, resulting in a mean $\mu=1.19$~meV, $\mu=-1.43$~meV, $\mu=-2.22$~meV and standard deviation per atom of $\sigma=3.36$~meV, $\sigma=8.94$~meV, $\sigma=11.44$~meV respectively clearly failing to fit the fat tailed distribution.  The green dashed line represents the Gaussian distribution obtained with the mean $\mu=0.72$~meV, $\mu=0.74$~meV, $\mu=-1.02$~meV and standard deviation per atom of $\sigma=6.85$~meV, $\sigma=19.92$~meV, $\sigma=22.13$~meV respectively corresponding to an error distribution with wider spread. ({\bf Bottom}): The histogram is constructed such the area is equal to unity. The black dashed line is a normalized Gaussian fit, resulting in a mean $\mu=0.012$~eV/\AA{}, $\mu=0.011$~eV/\AA{}, $\mu=0.002$~eV/\AA{} and standard deviation of $\sigma=0.08$~eV/\AA{}, $\sigma=0.10$~eV/\AA{}, $\sigma=0.12$~eV/\AA{} respectively clearly failing to fit the fat tailed distribution.  The green dashed line represents the Gaussian distribution obtained with the zero mean and standard deviation $\sigma=0.14$~eV/\AA{}, $\sigma=0.22$~eV/\AA{}, $\sigma=0.27$~eV/\AA{} respectively corresponding to an error distribution with wider spread }
\end{figure}


\section{Distance Analysis}

\begin{figure}[ht!]
    \centering
    \includegraphics[width=1.0\textwidth]{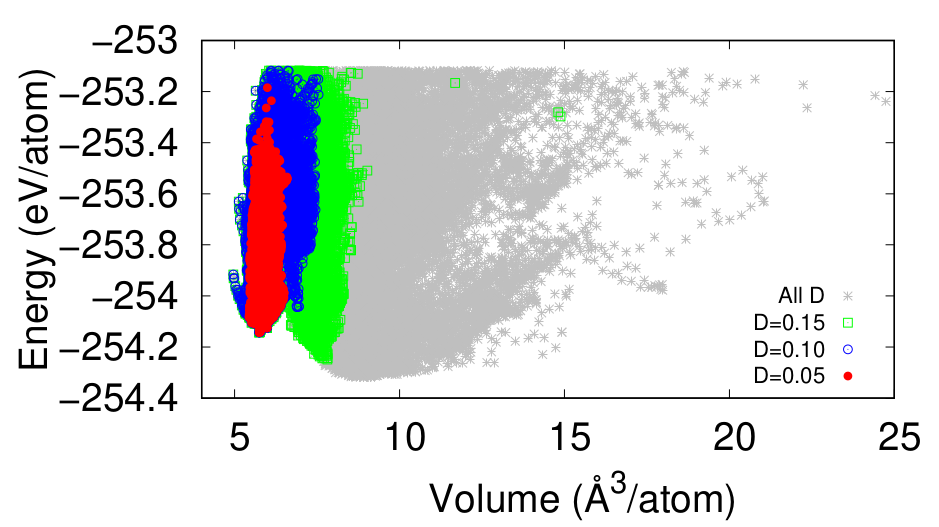}
    \label{fig:energy_volume_distance}
    \caption{Energy per atom as a function of volume for structures in the dataset, colored with respect to distance cutoff $D$ from diamond. }
\end{figure}

\begin{table}[hb!]
    \centering
    \caption{RMSE for energy in meV/atom (top) and forces in eV/\AA{} (bottom), for networks trained and validated on datasets of different distance to the reference phase, cubic diamond. T error is the average over the batch RMSE of the last 2500 training steps. The largest distance of any structure in the dataset to the reference phase is $D=0.62$.}
    \label{training_data_diversity}
\begin{ruledtabular}
\begin{tabular}{|l|c|c|c|c|c|c|}
\backslashbox{Train}{Validate}
&\makebox[6.5em]{T error }
&\makebox[6.5em]{All D }
&\makebox[6.5em]{$D=0.15$}
&\makebox[6.5em]{$D=0.10$}
&\makebox[6.5em]{$D=0.05$}
&\makebox[6.5em]{$D_{12}=0.05$}\\
\hline\hline
All D ($D=0.62$) & 22.0 
&~{\bf 22.1}& ~20.9      & 15.2      & 7.7  & ~~7.3 \\
$D=0.15$      & 18.1 
&  ~63.6    & ~{\bf 17.8}& 12.9      & 6.1  & ~16.9 \\
$D=0.10$      & ~8.7 
&  162.3    & ~52.2      & ~{\bf 9.4}& 4.3  & ~76.3 \\ 
$D=0.05$      & ~2.4 
&  473.8    & 219.0      & 75.3      &{\bf 2.5}& 474.2 \\
$D_{12}=0.05$ & ~2.6 
&  174.3    & ~88.3      & 52.0      & 2.7 & ~~{\bf 2.6}\\
\end{tabular}
%
\begin{tabular}{|l|c|c|c|c|c|c|}
\backslashbox{Train}{Validate}
&\makebox[6.5em]{T error }
&\makebox[6.5em]{All D }
&\makebox[6.5em]{\text{$D=0.15$}}
&\makebox[6.5em]{$D=0.10$}
&\makebox[6.5em]{$D=0.05$}
&\makebox[6.5em]{$D_{12}=0.05$}\\
\hline\hline
All D ($D=0.62$) &0.26 
& {\bf 0.27} & 0.27 &0.20 & 0.08 & 0.08 \\
$D=0.15$ & 0.24 
& 0.49 & {\bf 0.25} & 0.18 & 0.08 & 0.15\\
$D=0.10$ & 0.14 
& 0.96 & 0.44 & {\bf 0.15} & 0.06 & 0.31 \\
$D=0.05$ & 0.05 
& 1.67 & 1.05 &0.55 &{\bf 0.05} & 0.31 \\
$D_{12}=0.05$ & 0.05 
& 0.96 & 0.89 & 0.54 & 0.05 & {\bf 0.05} \\
\end{tabular}
\end{ruledtabular}
 \end{table}